\documentclass[aps,twocolumn,floatfix,groupedaddress,nofootinbib]{revtex4}
\expandafter\let\csname equation*\endcsname\relax
\expandafter\let\csname endequation*\endcsname\relax
\usepackage{csquotes}
\usepackage{graphicx,color}
\usepackage{float}
\usepackage[caption=false]{subfig}
\usepackage{amsmath}
\usepackage{amssymb}
\usepackage{multirow}

\begin{document}

\title{Dust remobilization in fusion plasmas under steady state conditions}

\author{P. Tolias,$^{1}$ S. Ratynskaia,$^{1}$ M. De Angeli,$^{2}$ G. De Temmerman,$^{3}$ D. Ripamonti,$^{4}$ G. Riva,$^{4}$ I. Bykov,$^{1}$  A. Shalpegin,$^{5,6}$ L. Vignitchouk,$^{1}$ F. Brochard,$^{5}$ K. Bystrov,$^{7}$ S. Bardin,$^{7}$ and A. Litnovsky$^{8}$}
\address{$^1$KTH Royal Institute of Technology, Association EUROfusion-VR, Stockholm, Sweden\\
             $^2$Istituto di Fisica del Plasma - Consiglio Nazionale delle Ricerche, Milan, Italy\\
             $^3$ITER Organization, Route de Vinon-sur-Verdon, CS 90 046, 13067 St-Paul-Lez-Durance Cedex, France\\
             $^4$Istituto per l'Energetica e le Interfasi - Consiglio Nazionale delle Ricerche, Milan, Italy\\
             $^5$Universit\'{e} de Lorraine, Institut Jean Lamour, Vandoeuvre-l\'{e}s-Nancy, France\\
             $^6$IGVP, Universit\"{a}t Stuttgart,  Stuttgart, Germany\\
             $^7$FOM Institute DIFFER, Dutch Institute For Fundamental Energy Research, Nieuwegein, The Netherlands\\
             $^8$Institute of Energy and Climate Research - Plasma Physics, Forschungszentrum J\"{u}lich, J\"{u}lich, Germany
}


\begin{abstract}
The first combined experimental and theoretical studies of dust remobilization by plasma forces are reported. The main theoretical aspects of remobilization in fusion devices under steady state conditions are analyzed. In particular, the dominant role of adhesive forces is highlighted and generic remobilization conditions - direct lift-up, sliding, rolling - are formulated. A novel experimental technique is proposed, based on controlled adhesion of dust grains on tungsten samples combined with detailed mapping of the dust deposition profile prior and post plasma exposure. Proof-of-principle experiments in the TEXTOR tokamak and the EXTRAP-T2R reversed-field pinch are presented. The versatile environment of the linear device Pilot-PSI allowed for experiments with different magnetic field topologies and varying plasma conditions that were complemented with camera observations.
\end{abstract}

\maketitle

\section{Introduction}\label{introductionsection}

\noindent Remobilization concerns the in-vessel release of dust residing away from its production site, whereas the general term mobilization also includes the release of dust upon creation (cracking, melt layer splashing, delamination, arcing). It has been long realised that dust remobilization is a major safety issue for ITER and future fusion devices, owing to the possibility of radioactive or toxic dust release upon loss of vacuum accidents (LOVAs)\,\cite{LOVAge1,LOVAge2,LOVAge3}. In such scenarios, air ingress in the vacuum vessel creates an outward flow after pressure equilibration which leads to hydrodynamic forces that can potentially mobilize dust grains\,\cite{LOVAth1,LOVAth2}. Moreover, dust remobilization upon disruptions has been consistently observed in multiple tokamaks by cameras\,\cite{Rudako1,Rudako2,ASDEXdu,MASTdeT,NSTXdet} and other diagnostics\,\cite{NSTXdet,HRTSobs}. In such cases, vibrations of plasma-facing components (PFCs), thermal shocks\,\cite{Martyne} or large currents induced by fast transients most probably provide the mobilizing forces. Finally, dust can also be remobilized in normal operating conditions by plasma forces. Regardless of the scenario, remobilization is a consequence of momentum imbalance in the \enquote{dust-PFC contact}. Hence, the main differentiating factor between remobilization in LOVAs, disruptions and normal operating conditions is the mobilizing force or torque.

Deeper understanding of the mechanism of remobilization by plasma forces can play an important role in diverse plasma-wall interaction issues.
\begin{itemize}
\item \emph{Specification of realistic initial conditions for the ejection speed and angle of remobilized dust grains}. This will increase the predictive power of dust transport codes as far as the grain penetration depth and the amount of dust-associated impurities are concerned. Consequently, this can lead to a more accurate modelling of transient impurity events\,\cite{TIEold1,TIEold2}, intense radiation spikes most likely associated with mobile dust re-distributed by a temporally adjacent disruptive discharge\,\cite{TIEnew1}.
\item \emph{Specification of realistic dust trajectory termination conditions}. So far, MIGRAINe is the only code that treats dust-wall interactions\,\cite{Mgrain1,Mgrain2,Mgrain3,Mgrain4} and thus has the potential to make predictions with respect to the in-vessel sites where dust will most likely accumulate. However, in the current version of MIGRAINe\,\cite{Mgrain2}, dust trajectories terminate for impacts that satisfy the sticking condition since remobilization is not accounted for. We stress that identification of preferred dust accumulation sites is the first step in developing efficient \emph{in situ} dust removal techniques. For instance, in ITER, mechanical dust removal is envisaged and hence successful predictions of the location of such sites can lead to dust collection by the divertor remote handling system or the multi-purpose deployment system\,\cite{ITERrhs}\,.
\item \emph{Quantification of dust amassment in the grooves of castellated PFCs}. In ITER, PFCs will be castellated, i.e. split into small segments separated by thin gaps\,\cite{Castel1,Castel2}. The gap entrance corresponds to a small fraction of the plasma-exposed area (the gap width is $0.5\,$mm), implying that dust trajectories directly terminating in the gaps can be considered rare. Two mechanisms related with the physics of the dust-PFC contact can be more efficient: dust remobilization from neighbouring monoblocks (provided that the release velocity is nearly tangential and low enough to ensure that the motion is governed not only by inertia but also by potential effects) or dust impacting and subsequently rolling or sliding on neighbouring monoblocks.
\item \emph{Evaluation of the gap trapping efficiency.} A closely related issue concerns grains already residing in the gaps and in particular whether they are permanently stuck therein or can be remobilized during normal plasma conditions or by a disruption.
\end{itemize}

Despite its importance, the remobilization of dust grains in normal operating conditions has not been properly considered thus far. There are no relevant experimental results, whereas previous theoretical investigations have either erroneously neglected the role of adhesion\,\cite{Remobi1,Remobi2,Remobi3,Remobi4} or adopted an oversimplifying description of the phenomenon\,\cite{Martyne}. In this work, we carry out the first combined experimental and theoretical study of dust remobilization in fusion plasmas during normal operating conditions. The main theoretical aspects of remobilization are analyzed, the role of adhesion in both establishing and breaking the dust-PFC contact is highlighted, whereas generic remobilization conditions are formulated. An experimental technique is proposed based on controlled pre-adhesion of dust grains on samples and detailed mapping of the deposition profiles prior to and post plasma exposure. Such a technique realistically mimics naturally occurring remobilization. Proof-of-principle of the technique is provided by experiments with planar tungsten samples in the TEXTOR tokamak and the EXTRAP-T2R reversed-field pinch. It is shown that plasma forces are sufficient to remobilize the grain and first quantitative estimates of the remobilization activity are performed. Further experiments in the Pilot-PSI linear device enabled studies under varying plasma conditions and magnetic field topology. They were complemented by camera observations allowing for estimates of the ejection speed and angle.

\section{Theoretical aspects of dust remobilization in fusion plasmas}\label{theorysection}

\noindent The theoretical analysis that follows is primarily focused on perfectly smooth surfaces and isolated spherical grains. The influence of non-ideal effects such as surface roughness and agglomerates is briefly discussed at the end of each subsection.

\subsection{The sticking impact regime}\label{stickingsubsection}

\noindent Before delving into the physics of remobilization, it is essential to discuss how dust gets originally stuck on plasma facing components. This is a crucial design element of controlled remobilization experiments, since, in order to allow for reliable extrapolations, dust grain deposition on samples needs to mimic dust sticking on PFCs as it naturally occurs in fusion environments. For a dust grain impinging obliquely on a smooth surface, when the normal impact velocity is smaller than a critical value known as the sticking velocity $v_{\mathrm{s}}\,$, all the normal kinetic energy of the grain is dissipated into adhesive work and plastic deformation. Consequently, the rebound velocity will be purely tangential and, in absence of strong external forces, it will slowly decrease owing to kinetic friction, until the dust grain is completely immobilized.

As has been discussed in Refs.\cite{Mgrain2,Mgrain3,Mgrain4,TNmodEx}, the typical impact duration ($\sim10\,$ns for the sizes of concern in this work) is sufficiently short to justify neglecting the work of plasma forces as compared to the impact energy of the dust grain and the total interface energy, \emph{i.e.} the work necessary to separate the two surfaces from contact to infinity. As we shall see in the following subsection, not only the work but also the normal component of the plasma forces themselves is negligible. Thus, while plasma defines the dust acceleration and bulk temperature (and hence the plastic dissipation upon collisions), the impact itself can be considered as purely mechanical. This opens up the way for a treatment of dust-PFC collisions with established impact mechanics models\,\cite{Stronge}. An analytical model of the normal component of elastic-perfectly plastic adhesive impacts originally developed by Thornton and Ning (T\&N model)\,\cite{Thorntn} has proved to be successful not only in reproducing experimental data obtained in vacuum\,\cite{TNmodEx} but also in explaining the general trends of experimental data obtained in plasma environments\,\cite{Mgrain4}. The T\&N model has been recently incorporated in the MIGRAINe dust dynamics code\,\cite{Mgrain1,Mgrain2,Mgrain3}. In the T\&N model, sticking is essentially controlled by two characteristic velocities, the adhesive velocity associated with adhesive work and the yield velocity associated with plastic work.

The \emph{adhesive velocity} is the maximum impact velocity for which an elastic-adhesive impact leads to zero rebound velocity. For such impacts, sticking is a consequence of the irreversible work of microscopic attractive inter-particle forces. Therefore, the adhesive velocity can be found by equating the normal impact kinetic energy of the grain with the total inelastic work carried out throughout the impact, from the initiation of the contact up to the separation of the surfaces of the colliding bodies. It naturally depends on the adhesive theory employed. Within the Johnson-Kendall-Roberts theory\,\cite{JKRtheo}, that is assumed in the T\&N model, it is given by
\begin{equation}
v_{\mathrm{s}}^{\mathrm{adh}}=\frac{\sqrt{3}}{2}\pi^{1/3}\sqrt{\frac{1+6\times2^{2/3}}{5}}\left(\frac{\Gamma^5}{\rho_{\mathrm{d}}^3E^{*2}R_{\mathrm{d}}^5}\right)^{1/6}\,,\label{adhesive}
\end{equation}
where $\Gamma$ is the interface energy per unit area, $\rho_{\mathrm{d}}$ is the dust mass density, $R_{\mathrm{d}}$ is the dust radius and $E^*$ is the reduced Young's modulus. The size dependence of the adhesive velocity is explicit.

The \emph{yield velocity} is the minimum impact velocity for which a pure elastic impact starts to become plastic. For the onset of plastic deformation beneath the contact region, the pressure at any point needs to exceed a limiting contact pressure $p_{\mathrm{y}}$ that is proportional to the yield strength $\sigma_{\mathrm{y}}$ of the material. For Hertzian profiles the maximum of the pressure is attained at the center of the contact area and the above criterion leads to a limiting contact radius $a_{\mathrm{y}}=\left(\pi{R}_{\mathrm{d}}p_{\mathrm{y}}\right)/\left(2E^*\right)$. Therefore, the yield velocity can be found by equating the normal impact kinetic energy of the grain with the elastic energy stored in the contact $U_{\mathrm{el}}(\delta_{\mathrm{y}})$ at a penetration depth $\delta_{\mathrm{y}}=a_{\mathrm{y}}^2/R_{\mathrm{d}}$. This leads to the expression
\begin{equation}
v_{\mathrm{y}}=\frac{\pi^2}{2\sqrt{10}}\left(\frac{p_{\mathrm{y}}^5}{\rho_{\mathrm{d}}E^{*4}}\right)^{1/2}\,.\label{yield}
\end{equation}
There is an implicit size dependence that is incorporated in $p_{\mathrm{y}}$. Material properties and especially the yield strength strongly depend not only on the bulk dust temperature $T_{\mathrm{d}}$ but also on the dust size. In fact, for metallic dust, as far as the onset of plastic deformation is concerned, there is a competition between the enhanced plasticity of heated metals and the high yield strength of micrometer-scale bodies. The $\Gamma(T_{\mathrm{d}})\,,\rho_{\mathrm{d}}(T_{\mathrm{d}})\,,E^{*}(T_{\mathrm{d}})$ and $\sigma_{\mathrm{y}}(T_{\mathrm{d}},R_{\mathrm{d}})$ dependencies for tungsten and beryllium have been described in Ref.\,\cite{Mgrain2}.

\begin{figure}
\centering
\includegraphics[width=3.4 in]{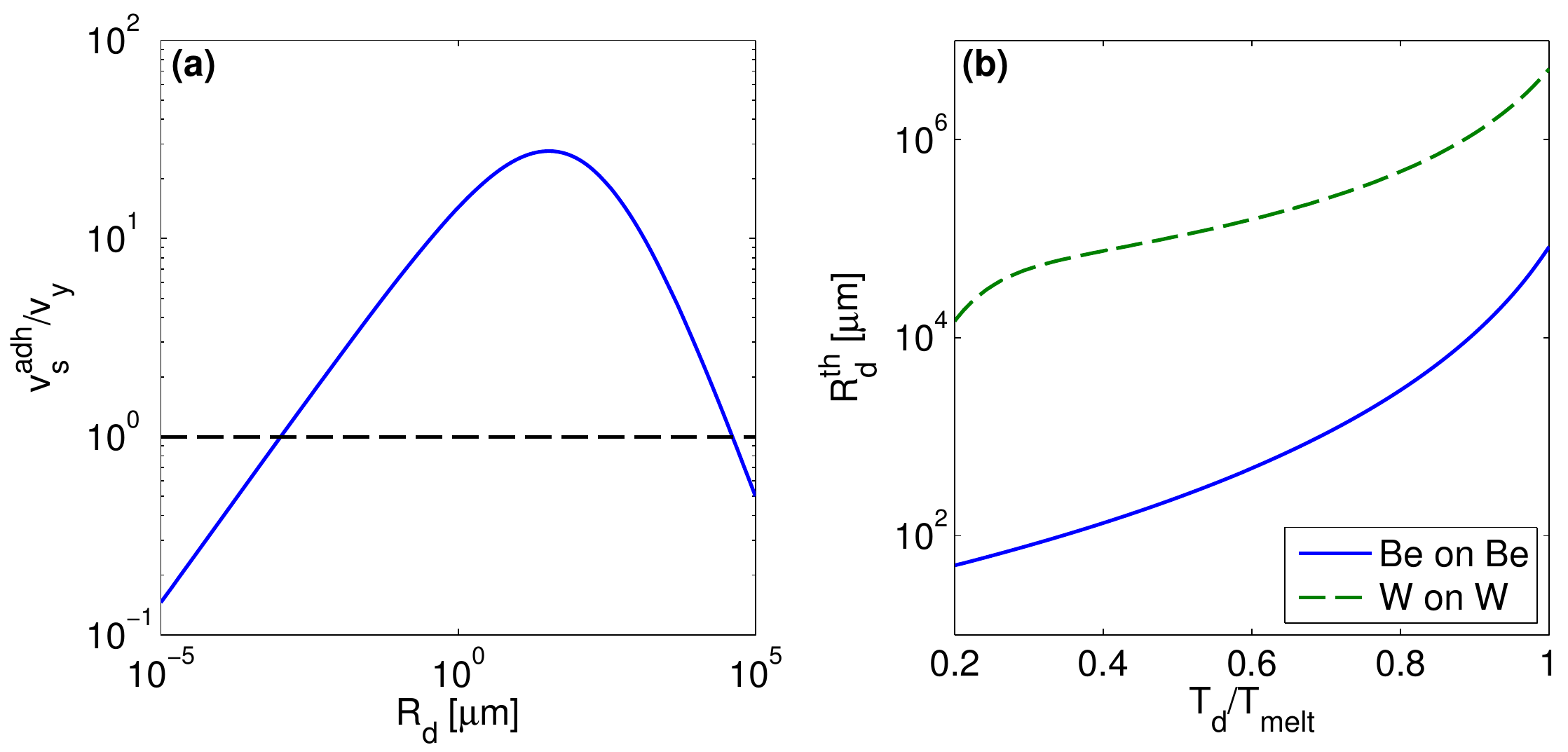}
\caption{(a) The ratio of the adhesive velocity to the yield velocity for W on W impact and $T_{\mathrm{d}}=1000\,K$. Notice that two solutions of the equation $v_{\mathrm{y}}(R_{\mathrm{d}})=v_{\mathrm{s}}^{\mathrm{adh}}(R_{\mathrm{d}})$ exist. For dust grains with radii that belong to the interval between the two solutions, $v_{\mathrm{y}}<v_{\mathrm{s}}^{\mathrm{adh}}$ and plastic deformation affects sticking. (b) Plot of the dust threshold radius as a function of the dust bulk temperature. For W dust impinging on W samples $R_{\mathrm{d}}^{\mathrm{th}}>10\,$mm for any temperature, while for Be dust impinging on Be samples $R_{\mathrm{d}}^{\mathrm{th}}\gg10\,\mu$m.}\label{plastic}
\end{figure}

Based on the value of these characteristic velocities, two distinct sticking realizations emerge. When $v_{\mathrm{y}}\geq v_{\mathrm{s}}^{\mathrm{adh}}$, plastic deformation does not occur during sticking impacts and the sticking velocity is equal to the adhesive velocity, $v_{\mathrm{s}}=v_{\mathrm{s}}^{\mathrm{adh}}$. On the other hand, when $v_{\mathrm{y}}<v_{\mathrm{s}}^{\mathrm{adh}}$, plastic work affects sticking and the sticking velocity is larger than the adhesive velocity, $v_{\mathrm{s}}>v_{\mathrm{s}}^{\mathrm{adh}}$. In order to distinguish between the two sticking realizations, it is convenient to define the dust threshold radius $R_{\mathrm{d}}^{\mathrm{th}}(T_{\mathrm{d}})$ that corresponds to the dust size for which the adhesive velocity and the yield velocity are equal. In Ref.\,\cite{Mgrain1}, the dependence of mechanical properties on $(T_{\mathrm{d}}\,,R_{\mathrm{d}})$ was not taken into account, which led to a single value of the dust threshold radius. Including these dependencies, we observe that for each value of the bulk dust temperature there exist two solutions of the equation $v_{\mathrm{y}}(R_{\mathrm{d}})=v_{\mathrm{s}}^{\mathrm{adh}}(R_{\mathrm{d}})$, see figure \ref{plastic}(a). The smaller solution is of no interest since it always lies in the nanometer scale; it is a manifestation of the Hall-Petch strengthening effect and demonstrates that for nano-sized dust plastic deformation is difficult to induce. Therefore, we shall identify the larger solution as the dust threshold radius $R_{\mathrm{d}}^{\mathrm{th}}(T_{\mathrm{d}})$: for $R_{\mathrm{d}}>R_{\mathrm{d}}^{\mathrm{th}}(T_{\mathrm{d}})$ plastic deformation has no effect in sticking.

The threshold radius is plotted in figure \ref{plastic}(b) as a function of the dust bulk temperature for W on W impacts and Be on Be impacts. We conclude that plastic deformation is always important for tokamak relevant impacts. Owing to the dependence of plastic deformation on the impact energy, after the dust grain is immobilized, the details of the contact region will also depend on the normal component of the grain impact velocity. This dependence on the pre-history of the trajectory clearly implies that control of the impact velocity prior to sticking should be an essential ingredient of reliable remobilization experiments. For tokamak relevant materials, the sticking velocity of $1-10\,\mu$m grains generally lies in the range $v_{\mathrm{s}}=0.1-10\,$m/sec\,\cite{Mgrain2}.

The above analysis of the sticking impact regime is valid for perfectly smooth spherical grains impinging on perfectly smooth planar surfaces. We stress that the omnipresent surface roughness can strongly reduce the sticking velocity, see Ref.\,\cite{Mgrain1} for a concise description. Finally, we point out that there is no theoretical description for non-spherical and non-planar colliding bodies.

\subsection{Simple estimates of the role of adhesion on remobilization}\label{adhesionsubsection}

\noindent In this subsection, we shall focus on the adhesive forces acting on the contact area that is established between the dust grain and the PFC after an impact in the sticking regime. For simplicity, we first assume perfectly smooth surfaces, spherical dust and planar PFCs as well as neglect plasticity. We shall analyze the elastic-adhesive contact with the aid of macroscopic contact mechanics.

We introduce the Hertzian elastic energy $U_{\mathrm{el}}$, the contact radius $a$ and the Hertzian penetration depth or depth of indentation $\delta=a^2/R_{\mathrm{d}}$\,\cite{Johnson}. A heuristic approach developed by Derjaguin\,\cite{Derjagu} assumes the validity of the Hertzian elastic contact equations and simply includes adhesion as a negative energy contribution\,\cite{FrenRev} approximated by $-\pi{a}^2\Gamma$. Therefore, the total energy of the system will be $U(\delta)=U_{\mathrm{el}}(\delta)-\pi{a}^2\Gamma$ and the contact force applied to the grain will be given by $F_{\mathrm{c}}(\delta)=\partial{U}/\partial{\delta}=F_{\mathrm{el}}(\delta)-\pi{R}_{\mathrm{d}}\Gamma\,$. Pull-off is achieved at $\delta=a=0$ and the pull-off force, \emph{i.e.} the minimum external normal force required to separate the surfaces, is given by $F_{\mathrm{po}}=\pi{R}_{\mathrm{d}}\Gamma$. Despite its simplicity, this heuristic model is useful for order of magnitude estimates. In fact, in more elaborate theories of elastic-adhesive contact, the pull-off force always has the form
\begin{equation}
F_{\mathrm{po}}=\xi_{\mathrm{a}}\pi{R}_{\mathrm{d}}\Gamma\,,\label{pulloff}
\end{equation}
with $\xi_{\mathrm{a}}$ a dimensionless coefficient of the order of a few\,\cite{GreenRe}. For instance, in the Johnson-Kendall-Roberts theory $\xi_{\mathrm{a}}=3/2$\,\cite{JKRtheo}, whereas in the Derjaguin-Muller-Toporov theory $\xi_{\mathrm{a}}=2$\,\cite{DMTtheo}.

The proportionality of the pull-off force with the dust radius highlights the dominant role of adhesion for micron-sized grains, since other forces acting on grains embedded in fusion devices scale either as $R_{\mathrm{d}}^2$ or as $R_{\mathrm{d}}^3$ (see below). A similar proportionality also stems from microscopic descriptions of the contact\,\cite{Hamaker}. Within the DLVO theory\,\cite{DLVOboo}, the total interaction between any two surfaces can be decomposed into the net Van der Waals force, as calculated with the additivity approach of Hamaker, and the electric double-layer force. For a sphere in contact with a flat surface, both contributions are proportional to the radius of the sphere.

We shall now compare the pull-off force with the main external forces; the ion drag force, the electrostatic force and the gravitational force. The discussion is restricted to the typical case of perfect electrical contact between the dust grain and the PFC, \emph{i.e.} in absence of insulating layers in the contact region. The well-known expressions for the drag and electrostatic forces\,\cite{Mgrain2,Lebedev} are not valid in our case, owing to the strong plasma inhomogeneity in the sheath and the effect of the boundary in binary collisions. Nevertheless, we shall employ these formulas for order of magnitude estimates, since more appropriate analytical expressions are not available in the literature. They are anyways expected to significantly overestimate the actual magnitude of the forces involved and will hence serve as upper limit estimates.

For a homogeneous infinite Maxwellian flowing plasma, the drag force due to the scattering of singly charged ions is given by\,\cite{IonDrag}
\begin{equation} F_{\mathrm{id}}^{\mathrm{sc}}=2\sqrt{2\pi}{R}_{\mathrm{d}}^2m_{\mathrm{i}}n_{\mathrm{i}}v_{\mathrm{Ti}}v_{\mathrm{i}}(z^2/\tau_{\mathrm{i}}^2)\ln{(\Lambda_{\mathrm{i}})}\mathcal{G}(u_{\mathrm{i}})\,,\label{iondrag}
\end{equation}
where $m_{\mathrm{i}}$ is the ion mass, $n_{\mathrm{i}}$ is the plasma density, $v_{\mathrm{Ti}}=\sqrt{T_\mathrm{i}/m_{\mathrm{i}}}$ is the ion thermal velocity, $v_{\mathrm{i}}$ is the ion flow velocity, $z=-e\phi_{\mathrm{d}}/T_{\mathrm{e}}$ is the normalized dust potential, $\tau_{\mathrm{i}}=T_{\mathrm{i}}/T_{\mathrm{e}}$ is the ion-to-electron temperature ratio, $\ln{(\Lambda_{\mathrm{i}})}$ is the effective Coulomb logarithm, $u_{\mathrm{i}}=v_{\mathrm{i}}/\sqrt{2}v_{\mathrm{Ti}}$ and $\mathcal{G}(u)=[\sqrt{\pi}\mathrm{erf}(u)-2ue^{-u^2}]/(2u^3)$. Similarly, for the drag force due to the absorption of ions we have $F_{\mathrm{id}}^{\mathrm{abs}}\propto{R}_{\mathrm{d}}^2$.

For a perfectly spherical conducting grain lying on a conducting plane in the presence of a uniform normal electric field $E_{\mathrm{w}}$, the boundary value problem for the determination of the electrostatic potential has an analytical solution\,\cite{Lebedev}. For such a configuration, the Laplace equation can be solved with the aid of degenerate bispherical coordinates and with one of the boundary conditions resulting from the constant behavior of the electric field at infinity. The contact charge of dust is given by $Q_{\mathrm{d}}=-\mathrm{\zeta(2)}R_{\mathrm{d}}^2E_{\mathrm{w}}\simeq-1.64R_{\mathrm{d}}^2E_{\mathrm{w}}$ and the normal electrostatic force acting on the grain is given by the formula $F_{\mathrm{E}}=\left[1/6+\mathrm{\zeta}(3)\right]R_{\mathrm{d}}^2E_{\mathrm{w}}^2$ or equivalently
\begin{equation}
F_{\mathrm{E}}=1.37R_{\mathrm{d}}^2E_{\mathrm{w}}^2\,,\label{lebedev}
\end{equation}
where $\mathrm{\zeta}(.)$ denotes Riemann's zeta-function. We shall also consider the gravitational force $F_{\mathrm{g}}=(4/3)\pi{R}_{\mathrm{d}}^3\rho_{\mathrm{d}}g\,$.

In the parameter range relevant to the ITER divertor conditions, we select favorable plasma conditions for the ion drag and electrostatic forces to be large, \emph{i.e.} $n_{\mathrm{e}}=10^{14}\,$cm$^{-3}$, $T_{\mathrm{e}}=T_{\mathrm{i}}=10\,$eV, $z=3$, $v_{\mathrm{i}}=\sqrt{2}v_{\mathrm{Ti}}$ and $E_{\mathrm{w}}=30\,$kV/cm. We also assume relatively large tungsten dust of $10$ micron diameter (several times larger than the median size of the tungsten grains collected in ASDEX Upgrade\,\cite{adhesi1}). Substituting the above parameters in Eqs.(\ref{pulloff},\ref{iondrag},\ref{lebedev}) we end up with
\begin{equation}
F_{\mathrm{po}}\sim10^2F_{\mathrm{id}}^{\mathrm{sc}}\sim10^3F_{\mathrm{id}}^{\mathrm{abs}}\sim10^3F_{\mathrm{E}}\sim10^6F_{\mathrm{g}}\,.\label{scaling}
\end{equation}
It is apparent that the pull-off force is orders of magnitude larger than the plasma forces and gravity. Therefore, previous investigations of dust remobilization that neglected adhesion are unrealistic\,\cite{Remobi1,Remobi2,Remobi3,Remobi4}.

Let us now consider the influence of non-ideal effects: \textbf{(i)} Plasticity will flatten out the grain increasing the contact area and thus the net adhesive force. In particular, permanent deformation of a spherical grain will lead to a local radius of curvature $R_{\mathrm{p}}>R_{\mathrm{d}}$ which will result to a stronger pull-off force of the form $F_{\mathrm{po}}\sim\xi_{\mathrm{a}}\pi{R}_{\mathrm{p}}\Gamma$\,\cite{Thorntn}. As mentioned in the previous subsection, $R_{\mathrm{p}}$ depends on the dust radius as well as the impact velocity. \textbf{(ii)} Non-spherical dust grains and non-planar PFCs will also affect the contact. For a qualitative estimate it is useful to consider Eq.(\ref{pulloff}) with the substitution $R_{\mathrm{d}}\to{R}_{\mathrm{eff}}$, where $R_{\mathrm{eff}}$ is the effective radius of curvature defined by $R_{\mathrm{eff}}^{-1}=R_{\mathrm{c,d}}^{-1}+R_{\mathrm{c,p}}^{-1}$ with $R_{\mathrm{c,d}}, R_{\mathrm{c,p}}$ the local radii of curvature of the grain and the PFC\,\cite{Johnson,GreenRe}. Apparently, PFC non-planarity will tend to decrease the pull-off force when the PFC can be approximated by a convex surface in the contact region (positive $R_{\mathrm{c,p}}$) and will tend to increase the pull-off force when the PFC can be approximated by a concave surface in the contact region (negative $R_{\mathrm{c,p}}$). Dust non-sphericity can also act both ways. We point out that, for non-spherical grains, the ion drag and the electrostatic force formulas also change, but generic expressions are not available in the literature. \textbf{(iii)} Nanometer scale roughness, \emph{i.e.} roughness consisting of asperities with dimensions much smaller than the size of the dust grain, is known to lead to a reduction of the contact area and consequently a reduction in the pull-off force only for stiff materials with large elastic moduli such as tungsten\,\cite{Mgrain1,TaborA1,TaborA2}. On the other hand, micrometer scale roughness, \emph{i.e.} roughness consisting of asperities with dimensions of the order of the dust size, is equivalent to PFC non-planarity and its effect on the pull-off force will also depend on the sign of the local radius of curvature. We note that roughness can also affect the plasma sheath structure and consequently the plasma forces, when the asperity heights become comparable to the Debye length. Such a scenario can be realized in partially detached divertor plasmas that are characterized by large plasma densities and small temperatures\,\cite{Stangeb}.

To sum up, non-ideal effects can either increase or decrease the pull-off force of Eq.(\ref{pulloff}), but they cannot account for the at least two orders of magnitude difference with the plasma forces, depicted in Eq.(\ref{scaling}). The only possible exception documented to be able to cause such a large decrease is nano-roughness\,\cite{TaborA2}.

Finally, it is important to make some remarks about the pull-off force between two spherical grains\,\cite{GreenRe}. This is relevant for the case of agglomerates, whose presence is unavoidable in our remobilization experiments. For spheres of radii $({R}_{\mathrm{d1}},{R}_{\mathrm{d2}})$, it is given by  $F_{\mathrm{po}}^{\mathrm{sp-sp}}=\xi_{\mathrm{a}}\pi\left[{R}_{\mathrm{d1}}{R}_{\mathrm{d2}}/({R}_{\mathrm{d1}}+{R}_{\mathrm{d2}})\right]\Gamma$, which results to
\begin{equation*}
\frac{F_{\mathrm{po}}^{\mathrm{sp-pl}}}{F_{\mathrm{po}}^{\mathrm{sp-sp}}}=1+\frac{{R}_{\mathrm{d1}}}{{R}_{\mathrm{d2}}}>1\,.
\end{equation*}
Therefore, not only adhesion is weaker in the sphere-sphere system than in the sphere-plane system but also the larger the size discrepancy the weaker the pull-off force. This is confirmed in our experiments, where it is systematically observed that agglomerates tend to break and remobilize easier than isolated grains.

\subsection{Force diagram for a dust grain in contact with the PFC}\label{diagramsubsection}

\noindent In our discussion of the sticking impact regime, we neglected the effect of plasma forces based on energy budget arguments. Moreover, in the previous subsection, we demonstrated that the pull-off force is orders of magnitude larger than plasma forces. However, plasma forces need to be responsible for remobilization, being essentially the only external forces exerted on the grain. In the formulation of remobilization conditions, due to the static nature of the problem, we can greatly benefit from the use of force diagrams. Very similar force diagrams are encountered in the field of aerosol science in industrial problems related to the hydrodynamic removal of contaminant dust\,\cite{aeroso1,aeroso2,aeroso3}. The main difference is that remobilization in those cases is induced by strong laminar or turbulent gas flows and not by plasma.

\begin{figure}
\centering
\includegraphics[width=3.2in]{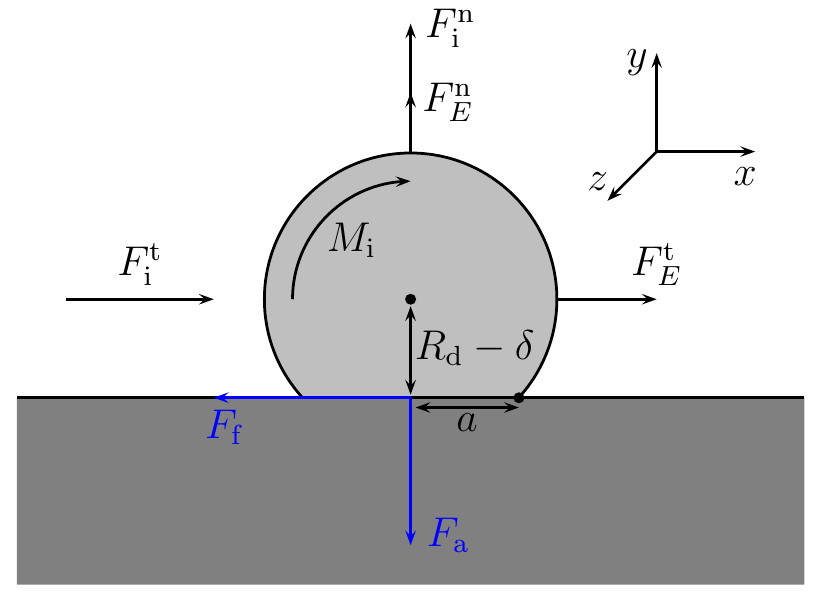}
\caption{Force diagram of a dust grain in contact with the PFC. \textbf{Note that the plasma forces carry a sign}. Both surfaces are assumed to be perfectly smooth, for simplicity it is also assumed that $\boldsymbol{F}_{\mathrm{i}}^{\mathrm{t}}\|\boldsymbol{F}_{\mathrm{E}}^{\mathrm{t}}$. The pull-off force $\boldsymbol{F}_{\mathrm{po}}$ is equal in magnitude and opposite in direction to the maximum value of the adhesive forces. The flattened and oversized contact radius is for pure illustrative purposes, in reality the shape of the contact area and of the deformed part of the sphere are complicated, with a so-called "neck" formation, see Ref.\cite{FrenRev} for details.}\label{forcediagram}
\end{figure}

The forces on a dust grain stuck on the PFC are sketched in figure \ref{forcediagram}. We neglect the gravitational force, since it is orders of magnitude smaller than the other forces. We also point out that for the typical case of small inclination angle of the magnetic field lines on the PFCs, there exist $\boldsymbol{E}\times\boldsymbol{B}$ and diamagnetic drifts arising in the pre-sheath and sheath regions\,\cite{sheath1,sheath2}. Therefore, as has been pointed out when analyzing the dust motion near a boundary\,\cite{sheath3}, ion flows of the same order exist in directions both parallel and perpendicular to the surface. This implies, also considering the flow direction not being an axis of symmetry and the strong inhomogeneity of the sheath, that plasma induced forces will have both tangential and normal components. Consequently, we have $\boldsymbol{F}_{\mathrm{i}}=\boldsymbol{F}_{\mathrm{i}}^{\mathrm{n}}+\boldsymbol{F}_{\mathrm{i}}^{\mathrm{t}}$ for the force due to the scattering and absorption of ions, $\boldsymbol{F}_{\mathrm{E}}=\boldsymbol{F}_{\mathrm{E}}^{\mathrm{n}}+\boldsymbol{F}_{\mathrm{E}}^{\mathrm{t}}$ for the force acting on the dust grain due to
the electrostatic field of the surrounding plasma sheath. Notice that the force $\boldsymbol{F}_{\mathrm{i}}$  can be assumed to have a line of action traversing the center of mass of the grain, provided that the excess torque is compensated for by the so-called external moment of surface stresses about the center, denoted here by $\boldsymbol{M}_{\mathrm{i}}$\,\cite{ExtMom1,ExtMom2}.

The other class of forces originates from the contact between the dust grain and the wall. In the normal direction, adhesive forces $\boldsymbol{F}_{\mathrm{a}}$ are exerted. They resist any external forces tending to separate the contact between the bodies. The minimum external force necessary to separate the two surfaces is the pull-off force $\boldsymbol{F}_{\mathrm{po}}$. In the tangential direction, frictional forces $\boldsymbol{F}_{\mathrm{f}}$ are exerted. They resist any external forces inducing relative motion between the bodies in contact. The maximum of the net frictional force, or equivalently the minimum external force necessary to cause sliding between the bodies is the static friction $\boldsymbol{F}_{\mathrm{fs}}$. We assume the validity of Amontons's law $F_{\mathrm{fs}}=\mu_{\mathrm{s}}F_{\mathrm{N}}$, with $\mu_{\mathrm{s}}$ the coefficient of static friction and ${F}_{\mathrm{N}}$ the normal component of the total force. It might seem contradictory to the reader that, even though we assume that both the dust grain and the wall are smooth surfaces, the coefficient of static friction is considered non-zero. In fact, in traditional theories\,\cite{Tribol1}, static friction is the consequence of surface asperity interlocking and arises from the dissipative processes that accompany the engagement / disengagement of micro-asperities. Such a picture is outdated\,\cite{Tribol2,Tribol3} and it has been experimentally demonstrated that even smooth surfaces can exhibit strong friction owing to atomic forces and local deformation processes. We note that, also in microscopic theories, static friction preserves its proportionality with the normal component of the force, in accordance with Amontons's law\,\cite{Tribol4}.

\subsection{Generic remobilization conditions}\label{conditionssubsection}

\noindent Assuming for simplicity that $\boldsymbol{F}_{\mathrm{i}}^{\mathrm{t}}\|\boldsymbol{F}_{\mathrm{E}}^{\mathrm{t}}$, the remobilization conditions can be found from force balance in the normal direction, force balance in the tangential direction and torque balance around the contact point\,\cite{Hydrod1,Hydrod2,Hydrod3}:
\begin{align}
&\mathrm{\underline{\emph{direct\,\,lift-up}}}:F_{\mathrm{i}}^{\mathrm{n}}+F_{\mathrm{E}}^{\mathrm{n}}>F_{\mathrm{po}}\,,\nonumber\\
&\mathrm{\underline{\emph{sliding}}}:F_{\mathrm{i}}^{\mathrm{t}}+F_{\mathrm{E}}^{\mathrm{t}}>\mu_{\mathrm{s}}\left(F_{\mathrm{po}}-F_{\mathrm{i}}^{\mathrm{n}}-F_{\mathrm{E}}^{\mathrm{n}}\right)\,,\label{remobconditions}\\
&\mathrm{\underline{\emph{rolling}}}:M_{\mathrm{i}}+(F_{\mathrm{i}}^{\mathrm{t}}+F_{\mathrm{E}}^{\mathrm{t}})(R_{\mathrm{d}}-\delta)>\left(F_{\mathrm{po}}-F_{\mathrm{i}}^{\mathrm{n}}-F_{\mathrm{E}}^{\mathrm{n}}\right)a\,.\nonumber
\end{align}
There is also the possibility of rotation around an axis parallel to the PFC surface normal (the $y-$axis in the coordinate system of Fig.\ref{forcediagram}). This is not considered in a detailed manner, since during pure self-rotation the position of the center of mass of the spherical grain relative to the PFC does not change.

Following Eq.(\ref{scaling}), it is clear that the direct lift-up condition cannot be easily realized in tokamaks under normal plasma conditions. On the other hand, owing to the small deformation due to adhesion $(a,\delta\ll{R}_{\mathrm{d}})$ the rolling condition is more easily realized. Finally, it is hard to draw any conclusions for the sliding condition due to the involvement of the coefficient of static friction, that is difficult to quantify.

Since the direct lift-up condition is hard to be satisfied in tokamaks, one could conclude that dust will remain on the wall once adhering to it. Nevertheless, the presence of micron-roughness, implies that while dust might initially roll or slide, it could eventually attain a velocity component normal to the local surface, \emph{i.e.} micron-roughness essentially acts as a ramp\,\cite{Mgrain1,Mgrain2}. Moreover, micron-roughness can also alter the form of the remobilization conditions, since it can lead to multiple contact points. For instance, in case roughness controls the point of rotation, the lever arms of both tangential and normal forces will change leading to a different rolling condition. This is also relevant for small agglomerates, see Refs.\cite{aeroso3,Hydrod3} for details. On the other hand, nano-roughness will not change the form of any of the remobilization conditions, it will only alter the numerical values of the pull-off force and the static friction coefficient.

\section{Experimental technique}\label{techniquesection}

\noindent The remobilization experiments are realized by controlled adhesion of micron-size dust on metal samples via gas-dynamic methods, exposure of the samples to plasma, and detailed mapping of the samples before and after their exposure.

\subsection{Controlled adhesion}\label{controlledsubsection}

\begin{figure}
\centering
\includegraphics[width=3.4 in]{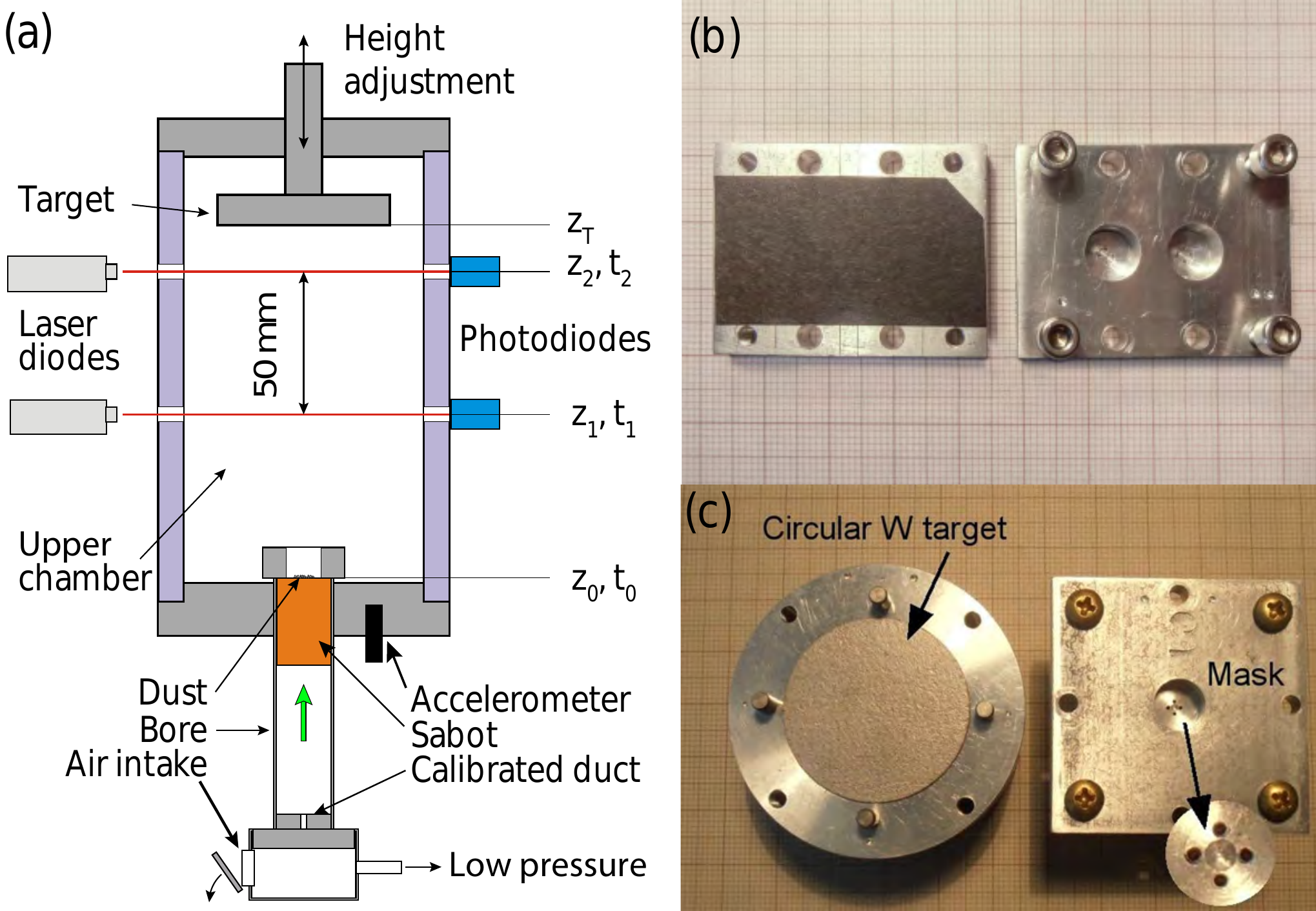}
\caption{(a) Schematic drawing of the modified pellet injection system. (b) Sample holder and mask for rectangular samples. (c) Sample holder and mask for circular samples.}\label{dust_launch}
\end{figure}

\noindent Controlled adhesion is achieved by a modified pellet injection system that launches dust grains with a pre-defined speed towards the metal surface. See figure \ref{dust_launch}(a) for a schematic drawing of the apparatus. Dust grains are deposited at the upper face of the sabot, whose initial position lies at the bottom end of the acceleration barrel. Low pressure $\simeq1\,$mbar is set into both the upper and the lower chamber, until the opening of the air intake, which induces an air flow that accelerates the sabot with a law depending on the size of a calibrated duct. The sabot impacts on the top of the acceleration barrel and ejects the dust grains into the upper chamber. The initial time of the dust grain release is identified by an accelerometer mounted on the bottom wall of the upper chamber that monitors the impact of the sabot. Ideally, the grain trajectory is purely deterministic, deceleration is caused by gravity and the initial velocity is equal to the sabot terminal speed. The grains follow a vertical path up to the metal target and, provided that their impact speed is lower than the sticking velocity, they adhere to the target.

The impact speeds that can be achieved with the apparatus range from $0.5\,$m/s up to $5\,$m/s. There are two factors that lead to uncertainties in the impact speed. Hydrodynamic forces associated with residual air in the upper chamber or with air leakages stemming from the lower chamber can slightly alter the force balance, whereas contact forces between the dust grains and the sabot can affect the initial velocity. Consequently, the dust velocity is monitored with two laser beams focused on the chamber axis. For each of the two laser diodes, three silicon-PIN-photodiodes (SIEMENS, SFH 229) are mounted on the wall, the one facing the laser diode ($10^4$ amplification factor) gathers the whole beam, while two additional sensors ($2.1\times10^7$ amplification factor) are placed at $\pm45^{\circ}$ from the beam direction in order to collect the scattered light. As the grains traverse the upper chamber and cross the laser beams, the attenuation of the signals is detected by the photodiodes facing the laser diodes, whereas the scattered light is detected by the other two photodiodes. The additional data provided by the second laser diode allow for a correction that accounts for grain interaction with air. We also point out that the vertical configuration of the system ensures that the grains remain immobile during the sabot acceleration.

A critical issue concerns the details of dust deposition at the sabot upper surface, which not only affect the total amount of released dust but can also lead to the undesired formation of agglomerates. This issue is of a great importance, particularly in the case of poly-disperse dust populations that include small size sub-populations of the order of $1\,\mu$m, since the latter have a strong tendency to form agglomerates when deposited on the sabot. A technique that reduces dust agglomeration is based on the use of glass spheres ($100-150\,\mu$m in diameter). The glass spheres are deposited on the sabot with their surface preemptively loaded with dust grains, which they release to the target upon impact. After releasing the grains, the glass spheres rebound and fall on the chamber bottom since their impact velocity $\sim1\,$m/s is much larger than their sticking velocity $\sim1\,$mm/s. We shall refer to this technique as mediated adhesion in order to differentiate from direct adhesion. We note, though, that formation of some small dust clusters on the sample is unavoidable.

\subsection{Dust deposition profile, sample mapping and surface roughness}\label{mappingsubsection}

\noindent In order to facilitate the pre- and post-exposure analysis of the samples and allow for an unambiguous identification of the dust spread and losses upon plasma exposure, it is convenient to confine the location of the adhered grains to well-defined spots, \emph{i.e.} circular large dust density spots with a sharp gradient at the edge. Such a control of the dust deposition profiles is achieved by the use of specially designed holders and masks. The samples are thus encased in machined holders and held in place by masks bearing circular holes of assigned size at specified locations. Two different holder / mask types were constructed, depending on the geometry of the samples, rectangular and circular. The rectangular mask bears two groups of four $0.5\,$mm diameter holes, lying on a $2\,$mm diameter circle, see figure \ref{dust_launch}(b). The circular mask has a single central group of four $0.5\,$mm diameter holes, placed symmetrically with respect to the center, see figure \ref{dust_launch}(c). In both cases, the thickness of the mask is decreased to $0.5\,$mm in the vicinity the groups in order to favour the dust passage.

The pre-deposited samples are mapped before and after plasma exposure, by means of a scanning electron microscope (SEM) with a $150$ magnification factor, adequate to resolve the smallest grains and sufficient to contain the $0.5\,$mm diameter dust spots in a single image. We note that a $400$ magnification factor (pixel size $0.25\,\mu$m) was used to collect images on the few spots where the grains are smaller than $2$ microns. The identification of remobilized grains was based on superimposing backscattered or secondary electron images before and after plasma exposure and implementing an image processing and analysis program, the NIH Image software\,\cite{softwar}. In order to allow for an absolute superposition of the images, specific mechanical reference marks were added to all the samples. The remobilization activity for each sample is presented by such overlaid images with specific color coding. The remobilized dust is generally separated in three different groups; grains that were removed from their initial position and whose terminal position is not clear (coded by red), grains that appeared after plasma exposure and whose initial position is not clear (coded by green), grains that were clearly displaced after plasma exposure with both their initial and terminal positions clear (coded by orange).

The roughness of the samples has been characterized by a surface profiler (KLA-Tencor P-15, KLA-Tencor Corporation - USA). It is quantified by the rms measure $R_{\mathrm{q}}$ defined as the root-mean-square vertical deviation of the roughness profile from the mean line measured in the sampling length. We shall also quantify roughness with the arithmetic mean measure $R_{\mathrm{a}}$, since it is typically used for roughness characterization in tokamaks\,\cite{roughn1,roughn2}. The measure $R_{\mathrm{a}}$ is defined as the mean absolute vertical deviation of the roughness profile from the mean line measured in the sampling length. The values of $R_{\mathrm{q}}$ and $R_{\mathrm{a}}$ are similar.

\subsection{Reference samples}\label{dummysubsection}

\noindent For all experimental campaigns, \enquote{reference} samples have been prepared. They have been treated in the same way as all other samples in terms of transportation, mounting on the device and vacuum exposure (pumping down and venting) but have not been exposed to the plasma. The reference samples are mapped before and after mounting in order to ensure that there is not any difference in the dust deposition profiles and hence that any kind of dust remobilization observed on the exposed samples is solely due to interaction with the plasma. Moreover, several tests on the strength of dust adhesion have been performed including for instance vibration in order to confirm that any accidental shaking of the sample will play no role in dust remobilization.

\section{Experimental results in fusion plasma environments}\label{experimentsection}

\noindent Experiments on dust remobilization under steady state conditions were carried out in the TEXTOR limiter tokamak, the EXTRAP-T2R reversed-field pinch and the Pilot-PSI linear device. The technical details and exposure statistics are summarized in Table \ref{summarytable}.

Adhesion is a purely deterministic phenomenon, \emph{i.e.} for a given impact speed and dust size, the pull-off force attains a unique value. However, the probabilistic nature of surface roughness and the uncontrollable occurrence of agglomerates, give adhesion a statistical character. Consequently, even pre-adhered dust spots exposed to identical plasma conditions can exhibit differences in their remobilization activity. Hence, a large number of experimental results is essential for quantitative assessments.

\begin{table*}
  \centering
  \caption{Remobilization experiments in fusion plasma environments.}\label{summarytable}
\begin{tabular}{c c c c}
\\
\multicolumn{1}{c}{\large{Device}}     & \multicolumn{1}{c}{\large{TEXTOR}}     & \multicolumn{1}{c}{\large{EXTRAP-T2R}} & \multicolumn{1}{c}{\large{Pilot-PSI}}             \\ \hline\hline
                                       &     \textbf{W rectangles}              &   \textbf{W squares}                   &  \textbf{W disks}                                 \\
\large{samples}                        &  $20\times40\,$mm$^2$, $t=0.2\,$mm     &   $8.5\times8.5\,$mm$^2$, $t=0.2\,$mm  &  $\varnothing=30\,$mm, $t=1\,$mm                  \\
                                       &  \,\,$R_{\mathrm{q}}=0.39\,\mu$m\,\,   & \,\,$R_{\mathrm{q}}=0.39\,\mu$m        & $R_{\mathrm{q}}=0.52\,\mu$m                       \\
                                       &  \,\,$R_{\mathrm{a}}=0.32\,\mu$m\,\,   & \,\,$R_{\mathrm{a}}=0.32\,\mu$m        & $R_{\mathrm{a}}=0.43\,\mu$m                       \\ \hline
                                       &     \textbf{spherical Ti}              &   \textbf{spherical W}                 &  \textbf{spherical W}                             \\
                                       &     $7-45\,\mu$m                       &   sphericity $>95\%$                   &  sphericity $>95\%$                               \\
\large{dust}                           &     Goodfellow                         &   $5-25\,\mu$m                         &  $5-25\,\mu$m                                     \\
                                       &     \textbf{irregular Mo}              &   purity $>99.9\%$                     &  purity $>99.9\%$                                 \\
                                       &     $<2\,\mu$m                         &   TEKMAT$^{\mathrm{TM}}$               &  TEKMAT$^{\mathrm{TM}}$                           \\
                                       &     Goodfellow                         &                                        &                                                   \\ \hline
                                       & \textbf{$\boldsymbol{1-2}\,$m/s}       &   \textbf{$\boldsymbol{1.0-1.7}\,$m/s} &  \textbf{$\boldsymbol{0.6-1.6}\,$m/s}             \\
\large{sticking}                       &  Ti, direct adhesion                   &   direct adhesion                      &  direct adhesion                                  \\
                                       &  Mo, mediated adhesion                 &                                        &                                                   \\ \hline
                                       &  \textbf{2 sets, 4 circular spots}     &  \textbf{4 circular spots}             &  \textbf{4 circular spots}                        \\
\large{spot}                           &  Ti, Mo sets                           &  symmetrically placed                  &  symmetrically placed                             \\
\large{geometry}                       &  $\varnothing_{\mathrm{spot}}=0.5\,$mm & $\varnothing_{\mathrm{spot}}=0.5\,$mm  & $\varnothing_{\mathrm{spot}}=0.5\,$mm             \\
                                       &  $\varnothing_{\mathrm{set}}=2\,$mm    & $\varnothing_{\mathrm{set}}=2\,$mm     &   $\varnothing_{\mathrm{set}}=2\,$mm              \\ \hline
                                       &     \textbf{32 spots}                  &   \textbf{12 spots}                    &  \textbf{40 spots}                                \\
\large{exposure}                       &     4 samples                          &   3 samples                            &  10 samples                                       \\
\,\,\,\,\large{statistics}\,\,\,\,     &     multiple discharges                &   single \&   multiple discharges      &  single discharges                                \\
                                       &                                        &                                        &  two re-exposures                                 \\ \hline
\large{$\measuredangle\boldsymbol{B}$} &    $\boldsymbol{18^{\circ}}$           &   $\boldsymbol{<5^{\circ}}$            &  $\boldsymbol{90^{\circ}}$ (5 samples)            \\
\large{to surface}                     &                                        &                                        &  $\boldsymbol{10^{\circ}}$ (5 samples)            \\ \hline\hline
\end{tabular}
\end{table*}

\begin{figure}
\centering
\includegraphics[width=2.9in]{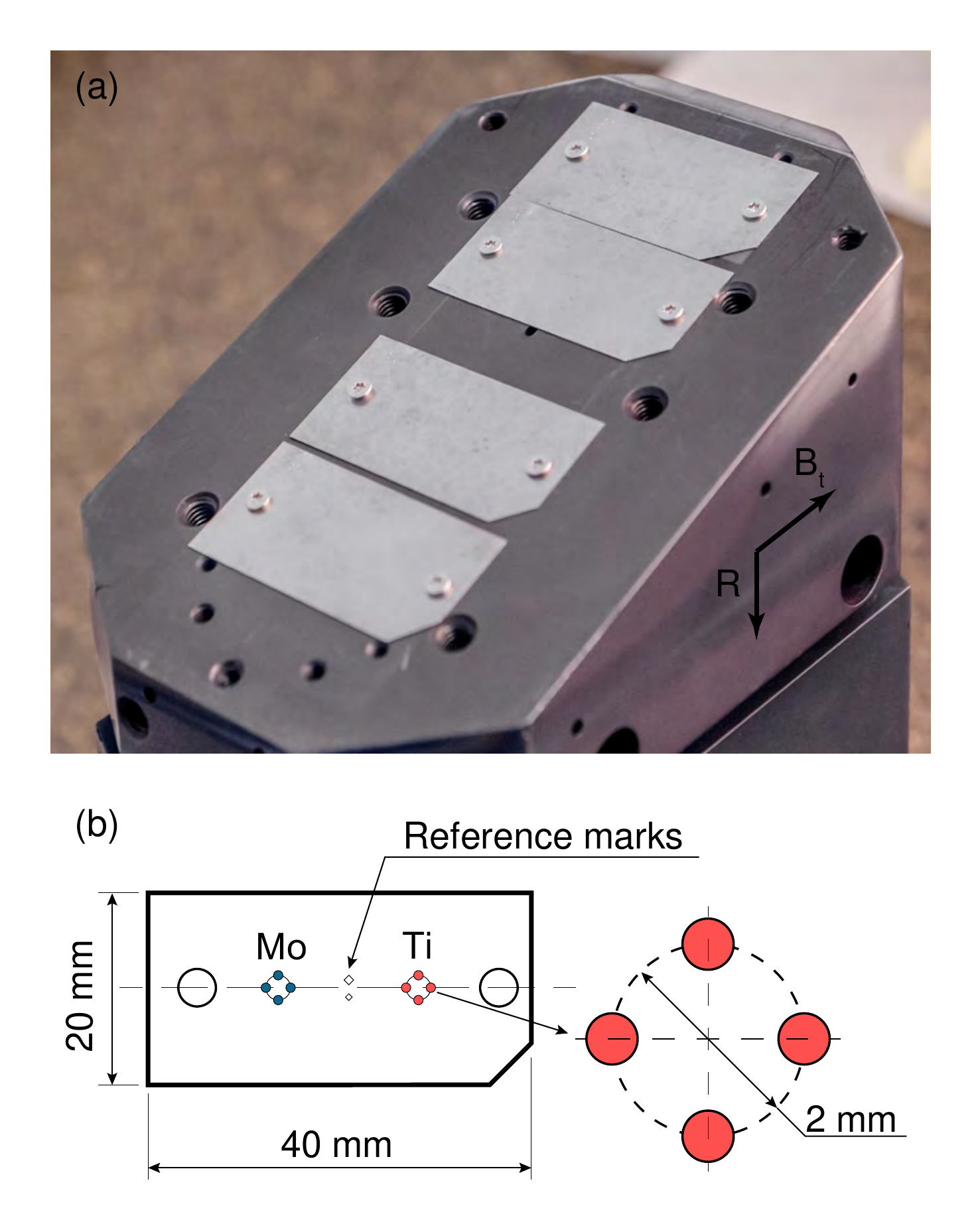}
\caption{(a) The tungsten samples mounted on LL1. The radial and toroidal directions are also visible. The limiter face has an inclination of 18$^{\circ}$ with respect to the toroidal magnetic field. (b) Layout of the tungsten sample and dust spot geometry. The plasma conditions on the Mo and Ti dust sets of the same sample are comparable.}\label{sa_TEXTOR}
\end{figure}

\subsection{Limiter tokamak TEXTOR}\label{TEXTORsubsection}

\noindent Four rectangular tungsten samples were simultaneously mounted on the TEXTOR bottom limiter lock I (LLI), see figure \ref{sa_TEXTOR}(a), that was inserted at a distance $R=52\,$cm from the torus center, \emph{i.e.}, $6\,$cm behind the last closed flux surface. The radial distance of the samples ranged from $52.4$ to $54.7\,$cm. The samples were exposed to discharges $\#120829$ and $\#120830$ for a total of $10\,$s ($6\,$s of plateau). Both discharges were NBI-heated with a total power of $1.0\,$MW, $2.25\,$T toroidal magnetic field, $350\,$kA plasma current and line-averaged plasma density of $3.5\times10^{19}\,$m$^{-3}$. The local plasma conditions at each sample differ, they vary in the range $n_{\mathrm{e}} = (1.2-2)\times10^{18}$\,m$^{-3}$, $T_{\mathrm{e}}=10-15\,$eV and $T_{\mathrm{i}}=3-8\,$eV\,\cite{Mgrain1} leading to a plasma Debye length $\lambda_{\mathrm{D}}\simeq10\,\mu$m that is comparable to the average size of titanium dust and much larger than the maximum size of molybdenum dust.

\begin{figure*}
         \centering\lineskip=-12pt
         \subfloat{\includegraphics[height=2.6in]{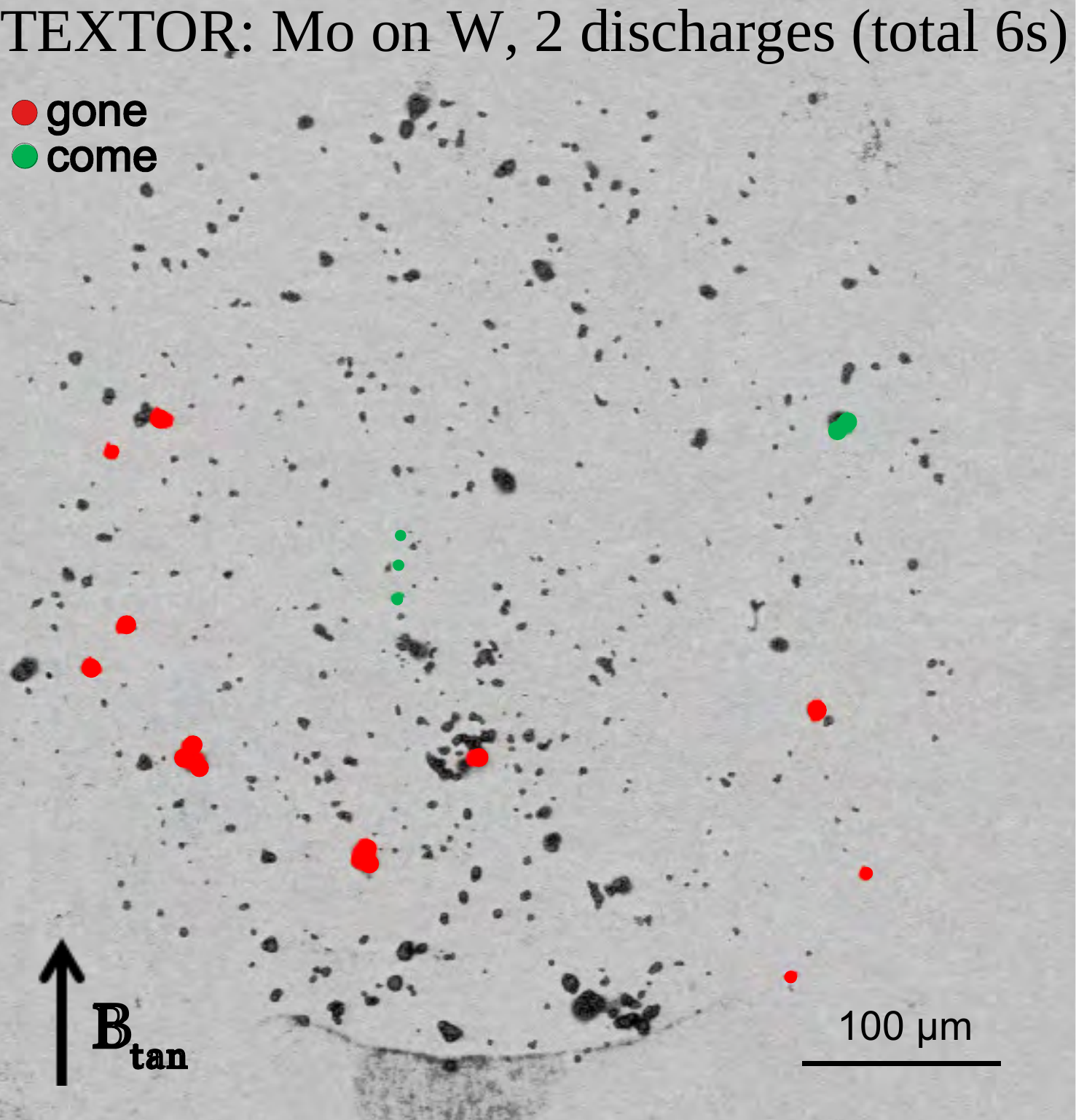}}
         \subfloat{\includegraphics[height=2.6in]{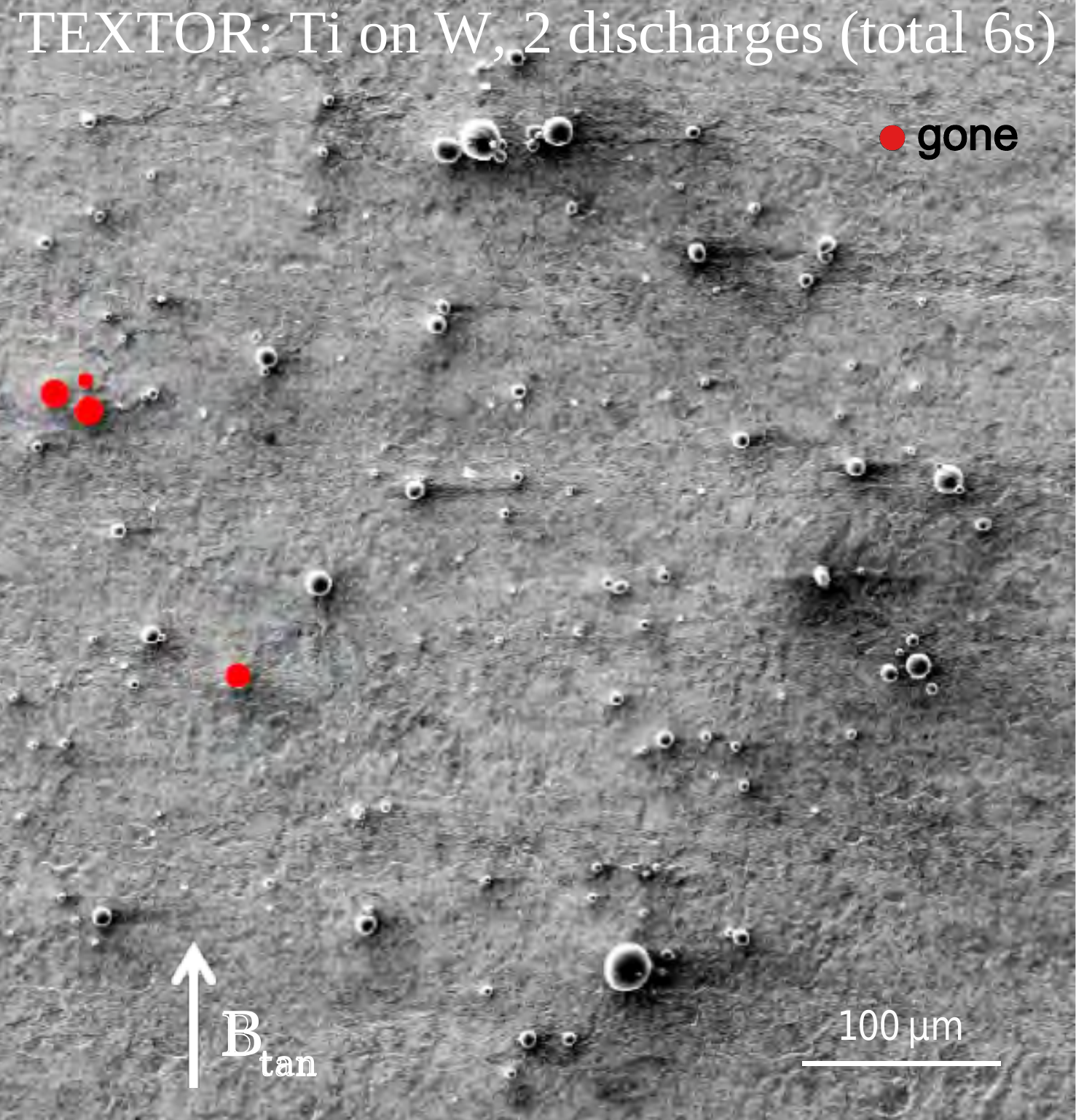}}
\caption{Overlaid backscattered electron image of one of the molybdenum dust spots of sample $\#1$ (left) and overlaid secondary electron image of one of the titanium dust spots of sample $\#2$ (right) before and after exposure to the TEXTOR plasma. The samples were exposed in two discharges for a total of $6\,$s of plateau. The direction of the tangential component of the local magnetic field is also indicated. Ten Mo agglomerates were removed from the spot, four Mo agglomerates were either displaced or originated from neighboring spots. Four Ti grains were removed from the spot, three with diameter around $10\,\mu$m and one with diameter around $3\,\mu$m. See also the online supplementary material, animations S1 and S2.}\label{moti_TEXTOR}
\end{figure*}

The TEXTOR experiment involved simultaneous tungsten dust injection experiments, hence grains of different refractory materials had to be employed for an unambiguous identification of the remobilization activity. Molybdenum dust of irregular shape with a nominal dimension $<2\,\mu$m and titanium dust of spherical shape with a nominal size distribution $7-45\,\mu$m was used. The spot geometry is shown in figure \ref{sa_TEXTOR}(b), each sample contains two dust spot sets of Mo and Ti. Each set has a $2\,$mm diameter and comprises of four symmetrically placed spots of $0.5\,$mm diameter. Since Mo grains have a strong tendency to agglomerate owing to their small size and large relative surface area, their adhesion to the samples was mediated by glass spheres. Mediated adhesion drastically reduced the size of the agglomerates, but a large amount of the adhered Mo dust was still in the form of clusters and a few agglomerates were still large with a surface area exceeding $200\,\mu$m$^2$. On the other hand, Ti grains were directly adhered. Both the Ti spheres and the glass spheres carrying Mo impacted the sample at a speed of $1-2\,$m/s.

\begin{table}
  \centering
  \caption{The remobilization activity for the TEXTOR experiments. The plasma conditions on each dust spot set are comparable, hence the remobilization activity is averaged over the four spots of each set for each sample, $\overline{\mathrm{RA}}=(1/4)\sum_{i=1}^{4}\mathrm{RA}_{i}$. Since the Mo dust shape was irregular and Mo was mostly adhered in the form of clusters, size separation in terms of surface covered was preferred. The dust surface was estimated by computing the apparent projected area in the backscattered electron images of the spots roughly accounting for extra contributions due to the presence of agglomerates. Notice that there was no remobilization activity in sample $\#3$.}\label{summaryTEXTOR}
\begin{tabular}{c c c c c}
$\boldsymbol{\overline{\mathrm{RA}}}$\,   & \textbf{Sample $1$}   & \textbf{Sample $2$}      & \textbf{Sample $3$}      & \textbf{Sample $4$}   \\
                                          & ($52.4\,$cm)          & ($53.1\,$cm)             & ($54.0\,$cm)             & ($54.7\,$cm)          \\ \hline\hline
Ti$>10\,\mu$m                             & $6.1\%$               & $4.2\%$                  & $0.0\%$                  & $5.6\%$               \\
Ti$<10\,\mu$m                             & $<0.1\%$              & $0.2\%$                  & $0.0\%$                  & $<0.1\%$              \\
Mo$>100\,\mu$m$^2$                        & $5.1\%$               & $0.5\%$                  & $0.0\%$                  & $0.0\%$               \\
Mo$<100\,\mu$m$^2$                        & $<1\%$                & $0.0\%$                  & $0.0\%$                  & $0.0\%$               \\ \hline\hline
\end{tabular}
\end{table}

Analysis of the samples has provided conclusive evidence of dust remobilization under normal tokamak operation conditions. We shall quantify remobilization by introducing the remobilization activity (RA), defined as the ratio of the dust grains within a selected size range that were removed or displaced during plasma exposure to the total number of grains (within the same size range) initially on the dust spot. The statistical analysis is presented in Table \ref{summaryTEXTOR} and overlaid SEM images of a Mo and a Ti dust spot are presented in figure \ref{moti_TEXTOR}. From the statistics available, total of 32 dust spots, the following picture emerges: \textbf{(i)} Spots with Mo agglomerates underwent nearly no changes with the exception of the sample deeper into the plasma ($R=52.4\,$cm), where $\simeq5.1\%$ of the largest agglomerates managed to remobilize, as shown in figure \ref{moti_TEXTOR}. \textbf{(ii)} Spots with Ti grains exhibited a more intense overall remobilization activity. Large grains $>10\,\mu$m tend to remobilize more than smaller grains $<10\,\mu$m, $4.0\%$ compared to $<0.1\%$. See figure \ref{moti_TEXTOR} for an example of moderate activity. \textbf{(iii)} We observed that few Ti grains or Mo clusters $<1\%$ were displaced from their original position without leaving the sample. \emph{Animations containing alternating SEM images of W samples prior and post exposure to the TEXTOR plasma are provided in the online supplementary material (see animations S1-S3)}.

\subsection{Reversed-field pinch EXTRAP-T2R}\label{T2Rsubsection}

\noindent In the reversed-field pinch EXTRAP-T2R\,\cite{Brunsel} of major radius $1.24\,$m and minor radius $0.183\,$m, three square tungsten samples were exposed, see also Table \ref{summarytable}. The exposure time of the samples varied from $20\,$ms (single discharge) to $330\,$ms (multiple discharges). The discharges were not identical, the plasma current was $70-100\,$kA, the plasma density $(0.5-1)\times10^{19}\,$m$^{-3}$ and the electron temperature in the range $100-200\,$eV. The samples were placed at the radius of the flat wall section, $11\,$mm outside the last closed flux surface, see figure 1 of Ref.\cite{adhesi2} for details of the vessel geometry. At this position the samples were shadowed by the main poloidal limiters at $0.183\,$m and by the bellows with inner radius $0.187\,$m. The surface normal was collinear with the vessel minor radius, thus the sample surface was tangential to the local, mainly poloidal, magnetic field. The local plasma parameters can be estimated as $n_{\mathrm{e}}\sim10^{17}\,$m$^{-3}$ and $T_{\mathrm{i}}\simeq{T}_{\mathrm{e}}\sim10\,$eV leading to a plasma Debye length $\lambda_{\mathrm{D}}\sim50\,\mu$m that is two times larger than the maximum tungsten dust size.

The sample geometry was similar to the TEXTOR experiments. Spherical tungsten dust grains TEKMAT$^{\mathrm{TM}}$ W-$25$ of $99.9\%$ purity were used, supplied by \enquote{TEKNA Advanced Materials Inc} with a nominal size distribution  $5-25\,\mu$m. SEM analysis confirmed the high sphericity of the grains but revealed a small percentage of irregular grains and the presence of a small sub-population with diameter below $5\,\mu$m. The W grains were directly adhered to the three W samples with impact velocities $1.6\pm0.1\,$m/s, $1.0\pm0.1\,$m/s and $1.7\pm0.1\,$m/s. The top of a few adhered W grains was marked by focused ion beam (FIB), such marks can potentially verify grain rotation during remobilization.

\begin{table}
  \centering
  \caption{The remobilization activity for the EXTRAP-T2R experiments. The plasma conditions on each dust spot set are comparable, hence the remobilization activity is averaged over the four spots of each sample, $\overline{\mathrm{RA}}=(1/4)\sum_{i=1}^{4}\mathrm{RA}_{i}$. In one of the spots of sample $\#2$, two large agglomerates remobilized that contained a large number of grains $<10\,\mu$m, thus biasing the value of $\overline{\mathrm{RA}}$ of this sample towards higher values. This spot is shown in figure \ref{w_EXTRAPT2R}.}\label{summaryEXTRAP}
\begin{tabular}{c c c c}
                                        & \textbf{Sample 1}             & \textbf{Sample 2}            & \textbf{Sample 3}            \\
$\boldsymbol{\overline{\mathrm{RA}}}$   & (1 discharge,                 & (5 discharges,               & (5 discharges,               \\
                                        & $\,\,\,\,v\simeq1.6\,$m/s)    & $\,\,\,\,v\simeq1.0\,$m/s)   & $\,\,\,\,v\simeq1.7\,$m/s)  \\ \hline\hline
W, all sizes                            & $9.1\%$                       & $17.1\%$                     & $6.5\%$                      \\
W, $>10\,\mu$m                          & $44.4\%$                      & $33.6\%$                     & $42.3\%$                     \\
W, $<10\,\mu$m                          & $7.9\%$                       & $16.3\%$                     & $4.4\%$                      \\ \hline\hline
\end{tabular}
\end{table}

\begin{figure}
\centering
\includegraphics[width=3.1in]{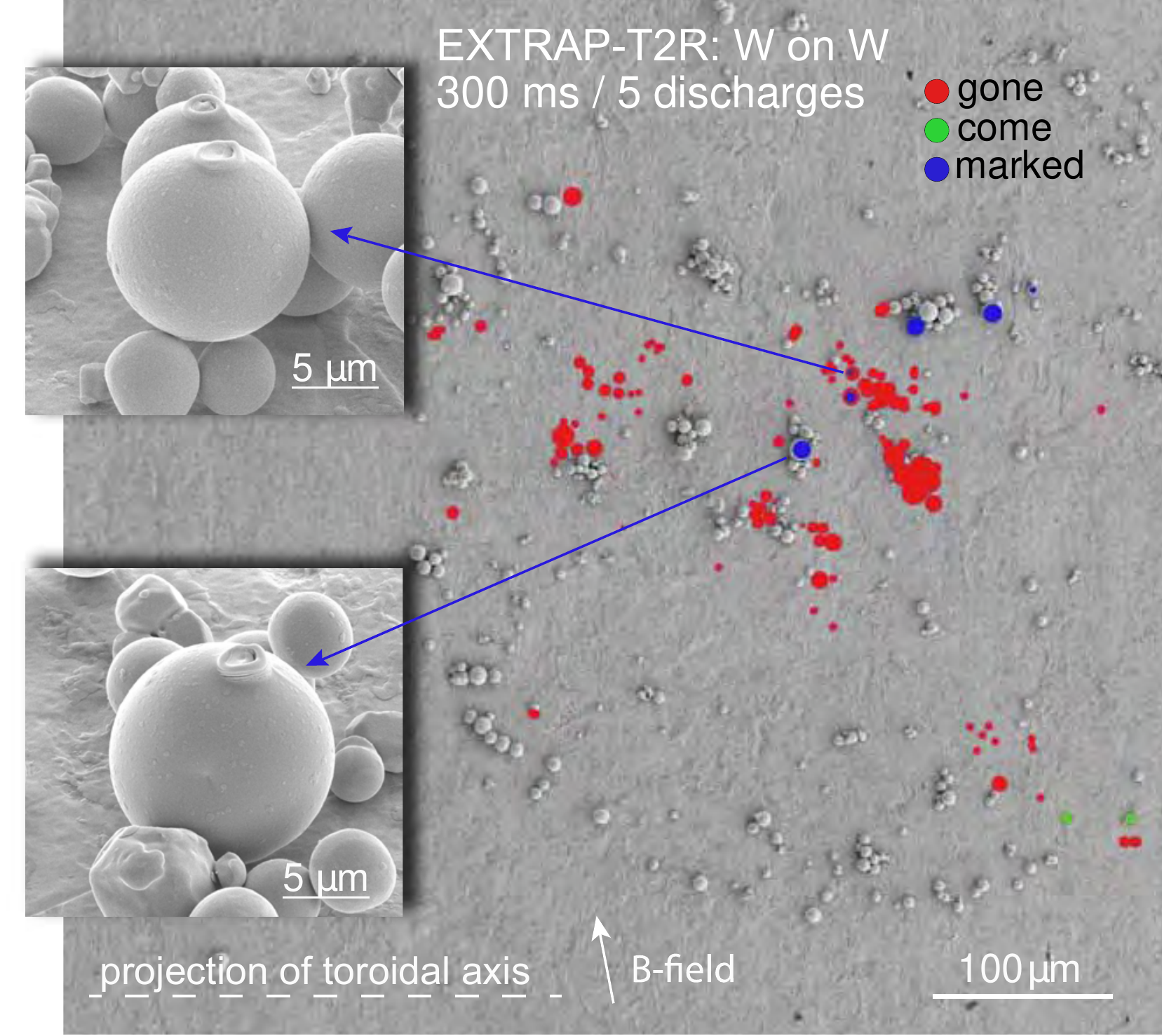}
\caption{Overlaid SEM image of one of the tungsten dust spots of sample $\#2$ before and after exposure to the EXTRAP-T2R plasma. The sample was exposed for $300\,$ms in 5 discharges. The direction of the nearly tangential local magnetic field is also indicated. A large number of isolated grains and agglomerates was removed from the spot; RA(total)$\simeq28.4\%$, RA($<10\,\mu$m)$\simeq28.1\%$, RA($>10\,\mu$m)$\simeq44.4\%$. Two isolated grains were either displaced or originated from neighboring spots. Two FIB-marked grains were removed, while the other four remained in their original position without exhibiting any signs of rotation. See also the online supplementary material, animation S4.}\label{w_EXTRAPT2R}
\end{figure}

Analysis of the samples has provided conclusive evidence of W dust remobilization. The statistical analysis is presented in Table \ref{summaryEXTRAP} and overlaid SEM images of a dust spot are presented in figure \ref{w_EXTRAPT2R}. Let us briefly summarize our results: \textbf{(i)} Large W grains $>10\,\mu$m remobilize easier than smaller grains $<10\,\mu$m, the respective remobilization activities averaged over all samples are $40.1\%$ compared to $9.5\%$. The difference in the remobilization activities is essentially much larger due to the fact that nearly $90\%$ of the remobilized smaller grains belonged to agglomerates. \textbf{(ii)} There are no appreciable differences between the sample that was exposed for a single discharge and the two samples that were exposed for multiple discharges. A similar average remobilization activity was observed, $\sim9.1\%$ and  $\sim11.8\%$, respectively. \textbf{(iii)} Very few tungsten grains $<0.1\%$ were only displaced during plasma exposure without leaving the sample. \textbf{(iv)} None of the FIB-marked grains that remained on the samples exhibited any signs of rotation. \emph{Animations containing alternating SEM images of W samples prior and post exposure to the EXTRAP-T2R plasma are provided in the online supplementary material (see animations S4-S6)}.

\subsection{Linear plasma device Pilot-PSI}\label{PSIsubsection}

\noindent \emph{Experimental set-up}. A large number of experiments were carried out in Pilot-PSI, see Table \ref{summarytable}, a linear plasma device capable of reproducing plasma conditions relevant for the divertor of ITER and future fusion devices\,\cite{PSIref1,PSIref2}. The hydrogen plasma is generated by a cascaded arc source that exhausts into a $40\,$cm diameter vacuum vessel and is confined by a strong axial magnetic field. The plasma column has Gaussian-shaped profiles with a full width at half maximum $\sim1\,$cm for the Pilot-PSI parameter space explored in our experiments.

\begin{figure}
\centering
\includegraphics[height=2.0in, width=2.2in]{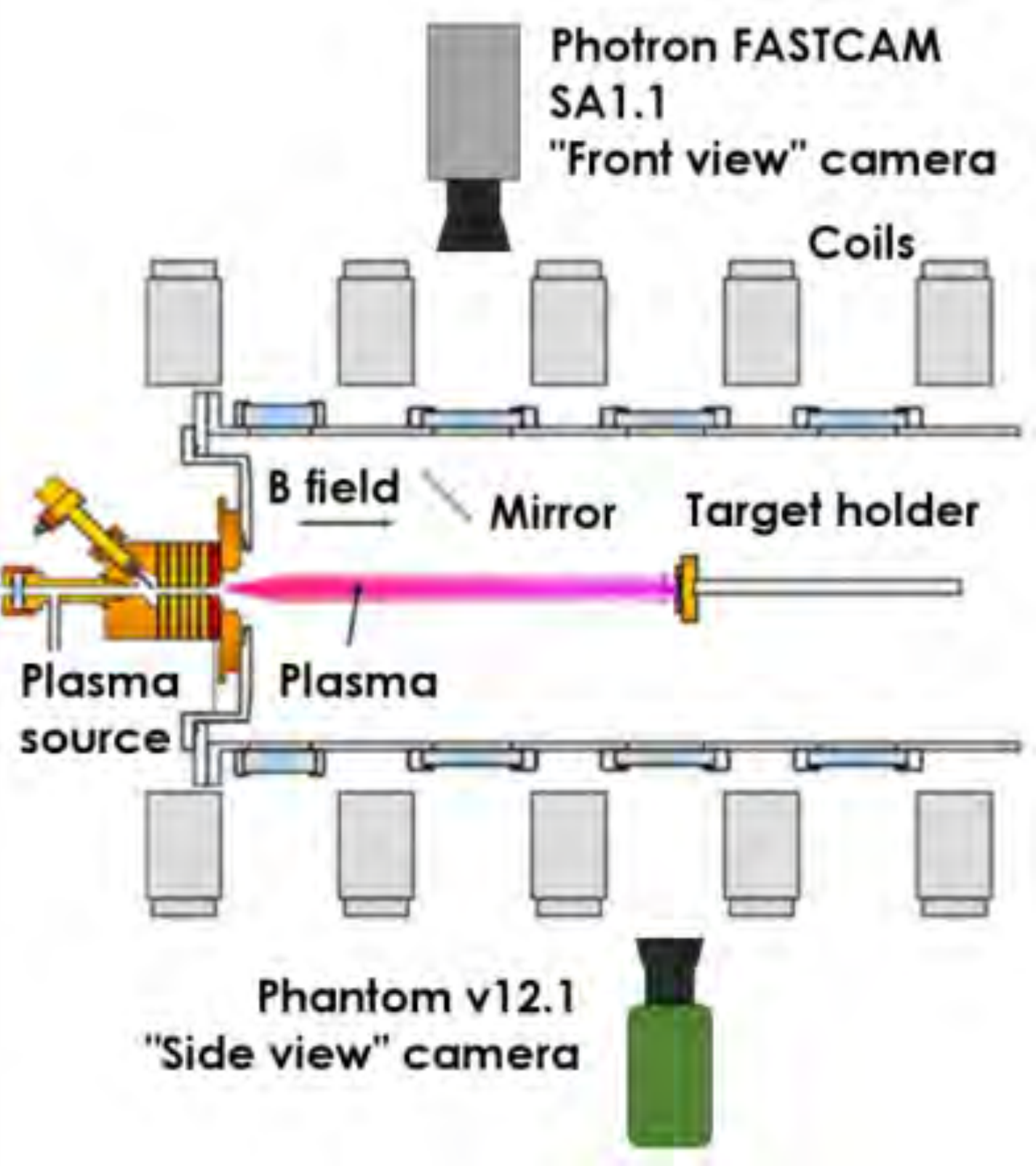}
\caption{Top view of the Pilot-PSI device with the camera arrangement. The target is concentric with the plasma column.}\label{sc_PSI}
\end{figure}

\begin{figure}
\centering
\includegraphics[width=3.0in]{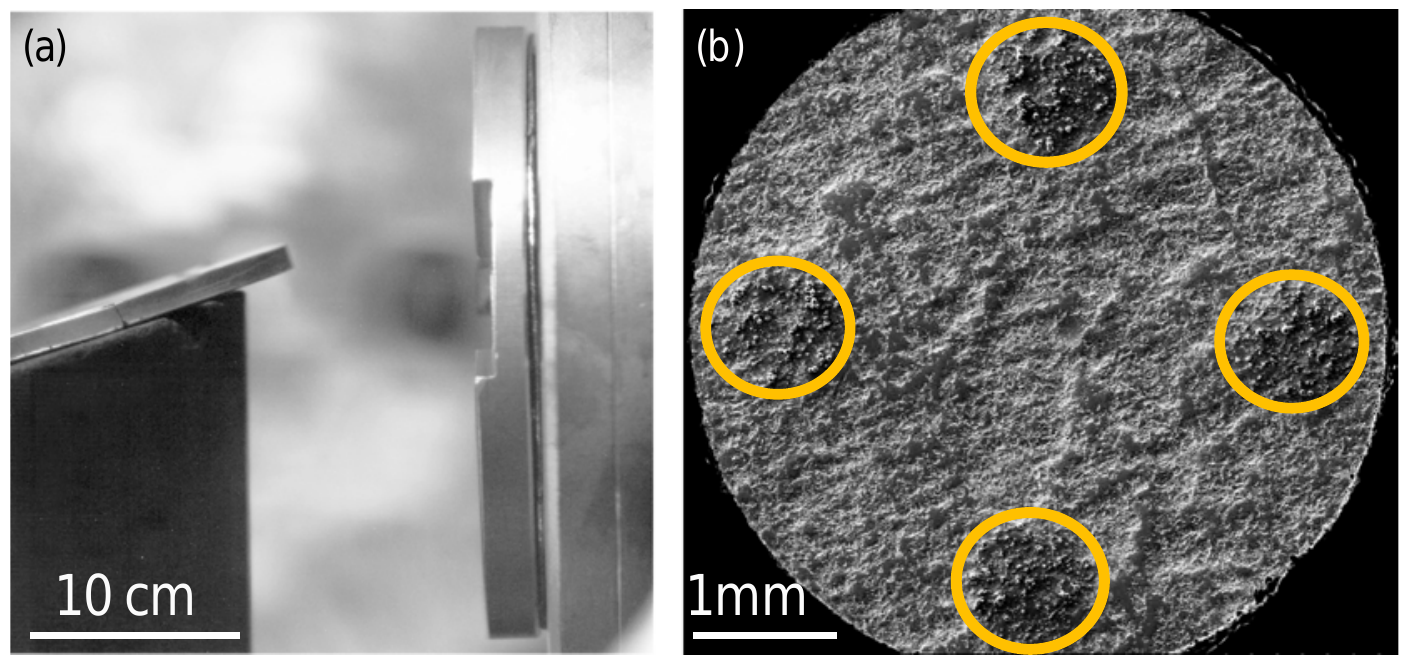}
\caption{(a) The endplate and the oblique plate inside the Pilot-PSI vessel. (b) SEM image of a dust spot set. The four symmetrically adhered dust spots are encircled.}\label{sa_PSI}
\end{figure}

The optimized diagnostic access of Pilot-PSI allowed for camera observations of remobilized particles and estimates of their ejection velocities. A schematic drawing of Pilot-PSI and the camera arrangement is presented in figure \ref{sc_PSI}. Two different set-ups were employed, see figure \ref{sa_PSI}(a). In the perpendicular configuration, the samples were mounted on the target at the end of the plasma column, henceforth referred to as endplate, with the magnetic field lines normal to the sample surface. In the oblique configuration, the samples were mounted on an inclined plate inserted at the vicinity of the endplate, henceforth referred to as oblique plate, with the magnetic field lines forming a $10^{\circ}$ angle with the sample surface. In all experiments, the endplate and the oblique plate were unbiased.

The operating and plasma parameters varied: magnetic field $0.4-0.8\,$T, discharge duration $1.5-2\,$s, plasma current $180-220\,$A, plasma density $(2-6)\times10^{20}\,$m$^{-3}$ and electron temperature $0.4-1.1\,$eV. For $T_{\mathrm{i}}\simeq{T}_{\mathrm{e}}$, the plasma Debye length was $\lambda_{\mathrm{D}}\simeq0.2\,\mu$m that is much smaller than the minimum tungsten dust size. These parameters refer to the plasma conditions at the endplate as measured by the Thomson scattering system. Such measurements could not be carried out for the oblique plate. Considering the geometry and size of the oblique obstacle but also plasma glow observations, we expect the local plasma to be more tenuous.

In these experiments, the W samples were disks of $30\,$mm diameter and $1\,$mm thickness, \emph{i.e.}, larger than the $1\,$cm plasma column. Each sample contains a single set of W dust spots, see figure \ref{sa_PSI}(b). The set has a $2\,$mm diameter and lies at the center of the disk (which coincides with the center of the plasma column), in order to ensure dense local plasma given the radial decay of the plasma profiles within the column. It comprises of four symmetrically placed W dust spots of $0.5\,$mm diameter. The W grains are TEKMAT$^{\mathrm{TM}}$ W-$25$ and were directly adhered to the W samples with impact velocities in the $0.6-1.6\,$m/s range. The dust density in the spots was intentionally larger relative to the previous experiments in order to increase the number of trajectories observed by cameras and thus permit a reliable estimate of the characteristic ejection speed and angle.

As in tokamaks\,\cite{Rudako1,tokdust}, visualization is achieved by capturing the thermal radiation emitted by dust, as a consequence of the high surface temperature that the released as well as the adhered grains reach by absorbing plasma fluxes. Trajectories of remobilized W grains were recorded with the aid of two high-speed cameras. The configuration of viewing ports allowed us to have two views on the W sample, perpendicular to each other, thus allowing for a stereoscopic reconstruction of the dust trajectories. A Photron FASTCAM SA1.1 camera (\enquote{front view}) was imaging the surface of the sample through a mirror installed inside the vacuum vessel, whereas a Phantom V12.1 camera (\enquote{side view}) was observing the edge of the sample directly through the window. Both cameras were synchronized with the Pilot-PSI triggering system and had similar recording parameters: a frame rate of $0.5-1\times10^4\,$frames/sec, a $1-20\,\mu$s exposure time depending on the experimental conditions and a $\sim200\,\mu$m/px spatial resolution in the image plane. In addition, the \enquote{side view} camera was equipped with one H\textrm{$\alpha$} band-stop filter to filter out excessive plasma light emission in all the experiments, while the \enquote{front view} camera was equipped with a H\textrm{$\beta$} band-stop filter only in the oblique configuration (one filter of each kind was available for these experiments). The filters allowed us to track grains for a longer time as well as to detect less bright grains.

The dust trajectories were reconstructed using the TRACE code\,\cite{TRACEc1,TRACEc2}. The videos were pre-processed in order to remove the background plasma light and maximize dust visibility. After background subtraction, nearly all light stemming from the plasma was removed from the frame, except from the high-contrast sample - plasma interaction region. The residual plasma light was removed by subtracting an averaged image composed of 10 neighboring frames. These techniques also allowed us to increase the frame contrast in the vicinity of the glowing spots of adhered dust and therefore trace dust grains closer to the point of their remobilization.

\emph{Perpendicular configuration}. Three samples were exposed to strong plasma ($n\sim6\times10^{20}\,$m$^{-3}$, $T_{\mathrm{i}}\simeq{T}_{\mathrm{e}}\simeq1.1\,$eV for $B=0.8\,$T). In the first, nearly all dust was removed from the sample, $\overline{\mathrm{RA}}(>10\,\mu$m)$=100\%$ and $\overline{\mathrm{RA}}(<10\,\mu$m)$\simeq98\%$. In the second, the majority of dust was either removed from the sample or strongly displaced destroying the circularity of the spots, $\overline{\mathrm{RA}}(>10\,\mu$m)$\simeq85\%$ and $\overline{\mathrm{RA}}(<10\,\mu$m)$\simeq35\%$. In the third, only grains belonging at the larger side of the size distribution exhibited high remobilization activity, $\overline{\mathrm{RA}}(>10\,\mu$m)$\simeq45\%$ and $\overline{\mathrm{RA}}(<10\,\mu$m)$\simeq5.5\%$. Notice that large variations arise in the values of $\overline{\mathrm{RA}}$ for each sample, in spite of the fact that all three samples were exposed to identical plasma conditions, the plates had the same average roughness characteristics and the impact velocities were the same. These large variations stem from the different sampling of the original dust population during the loading of the sabot that leads to different size distributions being adhered to the plates (see subsection \ref{controlledsubsection} for details). The sampling is not totally random, since different ways were employed to load the sabot that can favor different sizes. Thus, in few cases the sampling was observed to be strongly biased towards the largest dust sizes. The most biased case is that of the first sample. There, many adhered grains had diameters close to the maximum diameter present in the distribution, $25\,\mu$m, and formed agglomerates with smaller grains. On the contrary, in the third sample, even the largest adhered grains were typically smaller than $15\,\mu$m and most agglomerates consisted solely of smaller grains. The reader can observe these differences in the adhered populations by comparing animations S7 and S9 of the supplementary material. This is also supported by the fact that there are typically small variations in the values of $\mathrm{RA}$ for each spot of the same sample, since all spots were always shot simultaneously during a single dust loading of the sabot through the implementation of masks.

\begin{figure}
\centering
\includegraphics[height=2.8in]{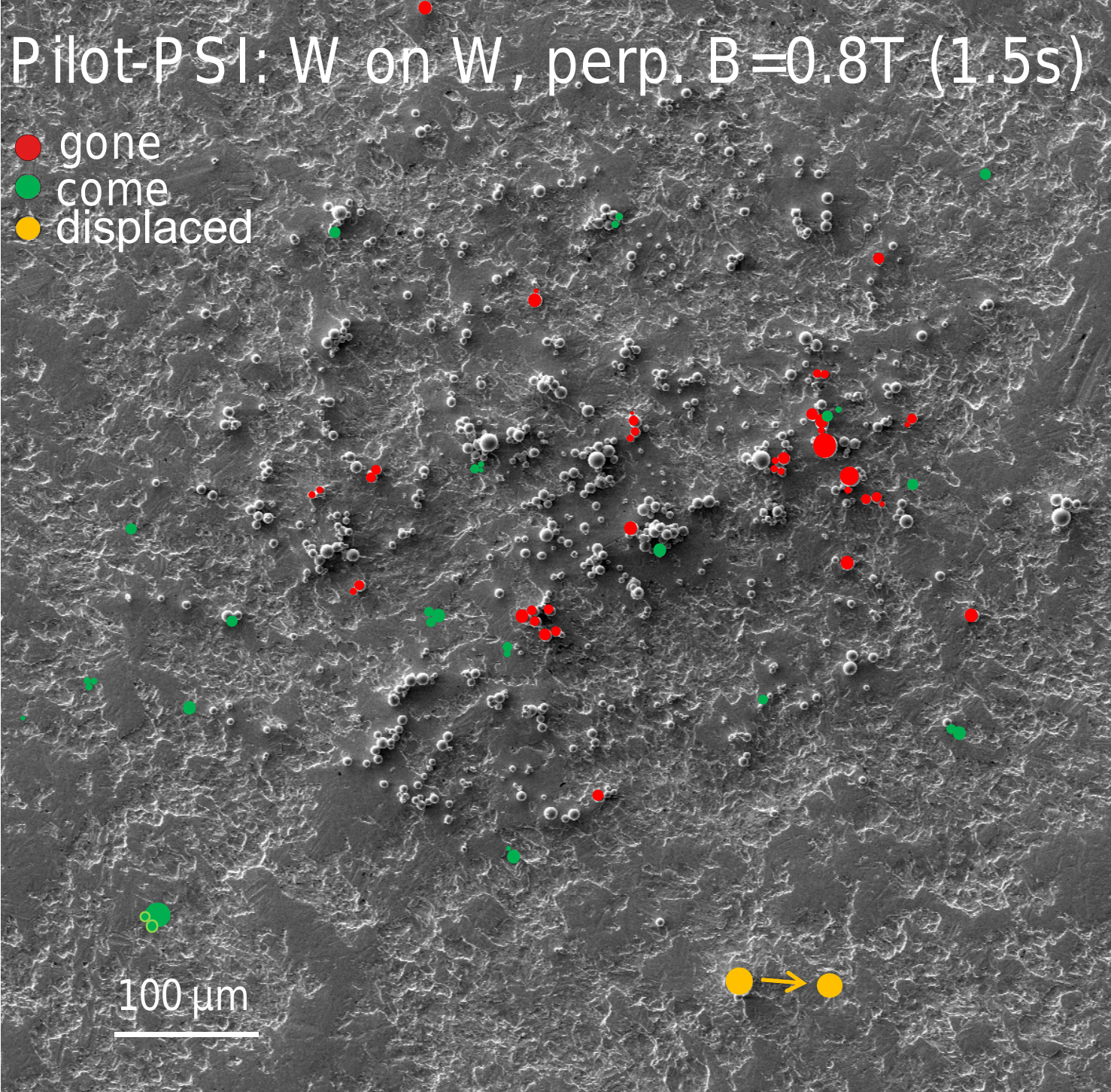}
\caption{Overlaid SEM image of a tungsten dust spot before and after exposure to the Pilot-PSI plasma. Exposure in the perpendicular configuration ($B=0.8\,$T, $t=1.5\,$s). The plasma parameters near the sample are $n\sim6\times10^{20}\,$m$^{-3}$ and $T_{\mathrm{e}}\sim1\,$eV. A large number of grains were either removed or strongly displaced (both isolated grains and agglomerates); RA($>10\,\mu$m)$\simeq50\%$, RA($<10\,\mu$m)$\simeq6\%$. We point out that the other two perpendicular exposures for $B=0.8\,T$ exhibited a much more intense remobilization activity; the spots were nearly removed, which makes overlaying SEM images redundant. See also the online supplementary material, animation S7.}\label{perp_PSI}
\end{figure}

Overlaid SEM images of a dust spot of the third sample are presented in figure \ref{perp_PSI}. Moreover, the second sample was re-exposed in identical plasma conditions. The re-exposure led to no additional remobilization. Finally, two samples were exposed to weaker plasma ($n\sim2\times10^{20}\,$m$^{-3}$, $T_{\mathrm{i}}\simeq{T}_{\mathrm{e}}\simeq0.4\,$eV for $B=0.4\,$T), nearly none of the adhered grains remobilized. In the 8 dust spots exposed, with $400$ being the average number of adhered grains on each spot, only four events were observed; one cluster consisting of two grains ($11\,\mu$m and $5\,\mu$m) was removed, one grain $\sim10\,\mu$m residing on top of a large agglomerate was removed, one large agglomerate containing a $20\,\mu$m grain slightly rotated, one agglomerate consisting of ten small grains $<8\,\mu$m weakly rotated. \emph{Animations containing alternating SEM images of W samples prior and post perpendicular exposure to the Pilot-PSI plasma are provided in the online supplementary material (see animations S7-S10)}.

For the case of strong plasma, released grains were recorded by cameras and a number of trajectories was reconstructed enabling a calculation of the average velocity. The released W grains move nearly parallel to the endplate, at a distance of $0.4\pm0.2$\,mm with an average speed of $\sim1.5\,$m/s. We did not manage to resolve the starting point of the trajectories. Typically, the initial observation point is a few millimeters away from the dust spot. This results from the combination of a number of factors: (i) The spatial resolution of the cameras is much larger than the size of the dust grains, $200\,\mu$m/pixel compared to $5-25\,\mu$m. Therefore, excessive plasma radiation is collected and only the hottest W grains can be observed. (ii) Isolated W grains are cooler when adhered, being in limited thermal contact with the cooler plate (limited due to the smallness of the contact area $\pi{a}^2$). Hence, in case remobilized grains initially slide or roll, their brightness might be insufficient. (iii) Thermal radiation from the adhered dust grains obscures camera vision close to the dust spots. The adhered grains are cooler but their cumulative radiation is strong due to their large number, $\sim400$ on average.

\begin{figure}
\centering
\includegraphics[width=3.4in]{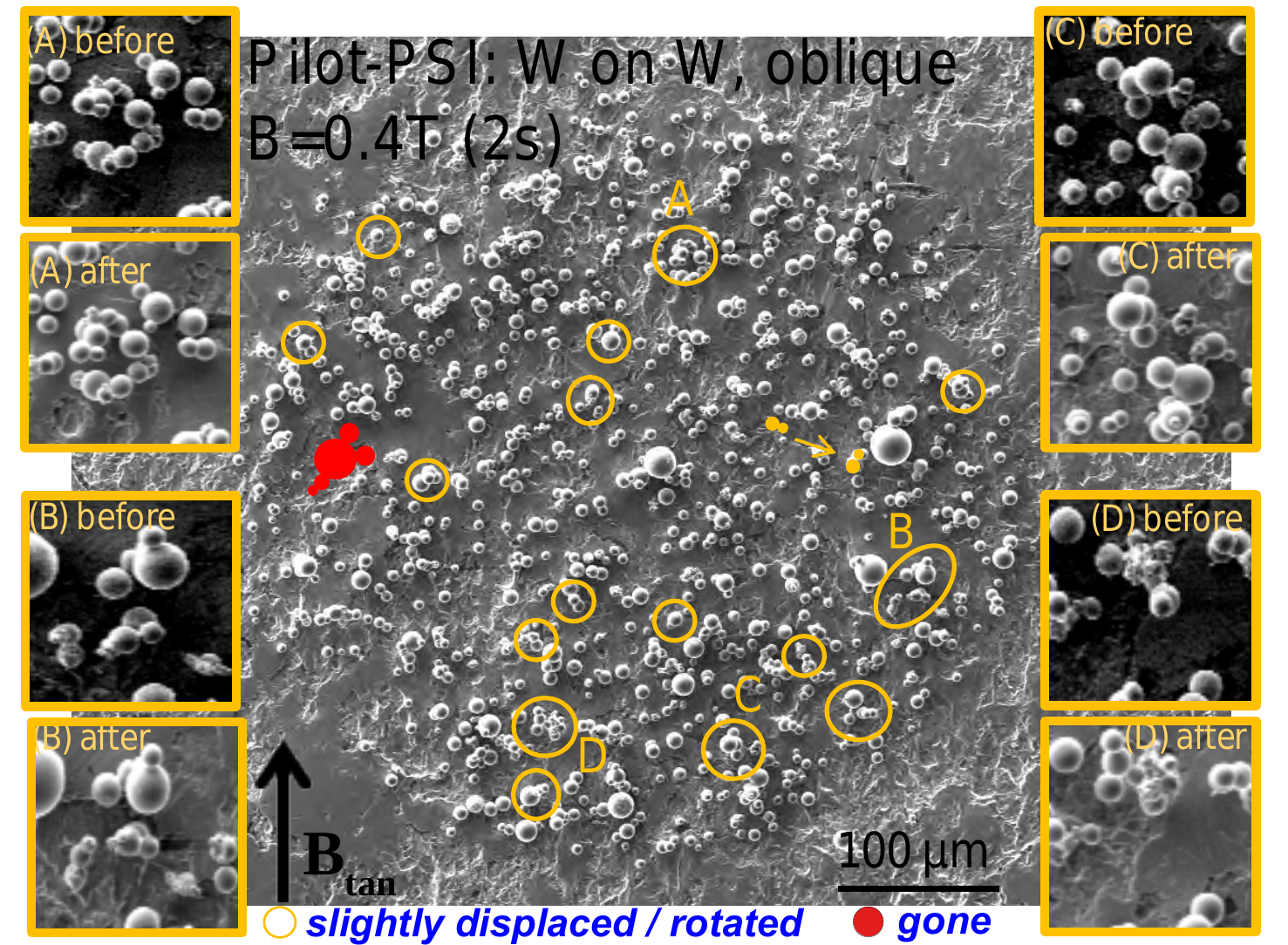}
\caption{Overlaid SEM image of a tungsten dust spot before and after exposure to the Pilot-PSI plasma. Exposure in the oblique configuration ($B=0.4\,$T, $t=2\,$s). The direction of the tangential component of the local magnetic field is also indicated. Only one W agglomerate was removed from the spot consisting of five dust grains with diameters $30,\,14,\,14,\,10$ and $6\,\mu$m. Two dust grains with diameters of $8\,\mu$m and $6\,\mu$m were strongly displaced. There are $16$ events of slight displacements or rotations (of the order of the dust size). See also the online supplementary material, animation S11.}\label{wind_PSI}
\end{figure}

\emph{Oblique configuration}. Five samples were exposed to plasma with $B=0.4-0.8\,$T. As aforementioned, in this configuration, plasma is expected to be much weaker on the sample due to the large size / small inclination angle of the perturbing oblique plate. All samples exhibited very low remobilization activity. In 9 dust spots no grains were removed. In the remaining 11 dust spots, with the number of adhered grains on each spot varying between $300-600$; there were 62 events where grains and small clusters were slightly displaced from their original position or weakly rotated with respect to their contact points (displacements are of the order of $10\,\mu$m, smaller than the average particle size), 30 grains were removed (18 have sizes larger than $10\,\mu$m, 12 have sizes smaller than $10\,\mu$m but only one of them was isolated), 6 grains were strongly displaced. Overlaid SEM images are shown in figure \ref{wind_PSI}. One of the samples was re-exposed in identical plasma conditions, the post and pre-exposure SEM images revealed no differences. \emph{Animations containing alternating SEM images of W samples prior and post oblique exposure to the Pilot-PSI plasma are provided in the online supplementary material (see animations S11,S12)}.

\begin{figure}
\centering
\includegraphics[width=3.2in]{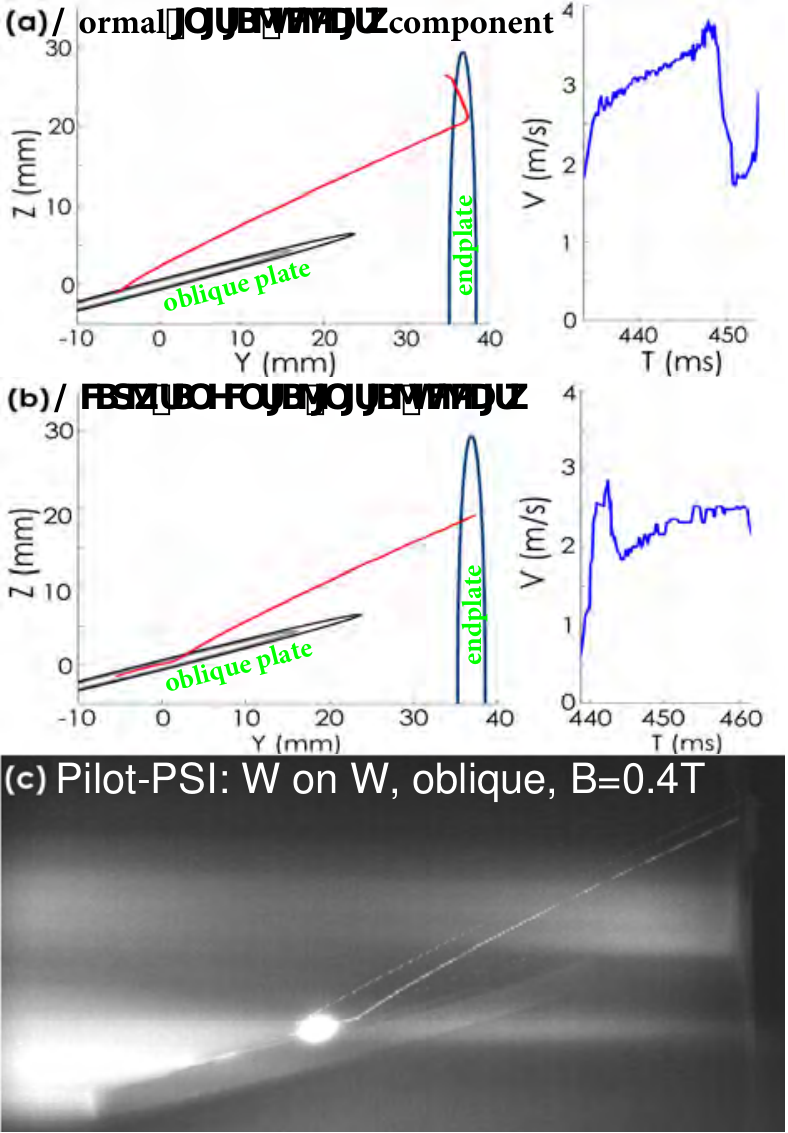}
\caption{Camera observation of remobilized W dust in Pilot-PSI. Exposure in the oblique configuration ($B=0.4\,$T, $t=2\,$s). Notice that for both cases the remobilization instant is $t_{\mathrm{rem}}<440\,$ms, see also discussion in section \ref{summarysection}. \textbf{(a)} The trajectory of a remobilized W grain with an initial normal (to the oblique plate) velocity component and the calculated total speed as a function of time. The grain is accelerated by the ion drag force until it collides with the endplate. The impact is inelastic and part of the impact kinetic energy is lost mainly due to plastic deformation and the adhesive work\,\cite{Mgrain1,Mgrain2,Mgrain4}. \textbf{(b)} The trajectory of a remobilized W grain with an initial nearly tangential (to the oblique plate) velocity and the calculated total speed as a function of time. Shortly after the beginning of observation, there is a brief interval where the grain's kinetic energy is reduced, $t=443\,$ms up to $t=445\,$ms. Such a deceleration can either be caused by friction (in case the grain moves along the oblique surface) or by the competition between the electrostatic sheath force and the ion drag force (in case the grain moves above but close to the oblique surface). However, due to the lack of sufficient resolution, other scenarios such as dust multi-bouncing cannot be excluded\,\cite{Mgrain4}. After its detachment, the grain is accelerated by the ion drag force until it collides with the endplate. The normal (to the endplate) velocity is below the sticking velocity and the grain remains adhered to the endplate, where it rapidly cools down and its brightness falls below the camera threshold. \textbf{(c)} Superposition of frames from camera observations depicting the two trajectories. It is evident from the brightness of the column that the plasma is quenched on the beginning of the oblique plate, before reaching the sample (the dust set corresponds to the bright spot), which is mounted in the middle.}\label{oblique_PSI}
\end{figure}

Also for the oblique configuration, a number of trajectories was reconstructed by the camera observations. Some trajectories had an initial velocity component normal to the oblique plate, see figure \ref{oblique_PSI}(a), while most trajectories were initially nearly tangential to the oblique plate, see figure \ref{oblique_PSI}(b). The resolution is not sufficient to allow us to determine whether these grains are moving along the surface of the oblique plate or along the plasma sheath, but we have confirmed that their tangential motion is primarily along the plasma stream. As in the perpendicular configuration, the grains become visible after the onset of remobilization, hence it is not possible to determine whether trajectories of the first type, initially had a purely tangential character.

\section{Discussion and conclusions}\label{summarysection}

\noindent The present studies refer exclusively to metallic dust remobilization from metallic surfaces under normal operating conditions (in absence of transient events) with the grain motion triggered by the action of plasma forces. As will be discussed in details below, camera observations in Pilot-PSI clearly illustrate that dust remobilization takes place \emph{during plasma exposure}. This is further supported by the reference samples not exposed to plasma (see also subsection \ref{dummysubsection}), which never exhibited any remobilization activity.

Our TEXTOR and EXTRAP-T2R experimental results confirmed that remobilization does take place under steady state plasma operation and revealed that the condition for the grain release is not easily satisfied, since only a relative small fraction of the exposed populations has remobilized and among those mostly larger grains and small clusters. The interpretation of the Pilot-PSI data greatly benefits from being complemented by camera observations and from the multiple exposure of samples to identical plasma conditions. The experimental evidence from this device allow us to draw the following rather important conclusion:  For a given \enquote{dust-PFC contact}, there appears to be a condition to satisfy for remobilization to occur, in case it is not fulfilled in the beginning of the plasma exposure, then the grain shall remain adhered provided that the plasma conditions are stationary.

Let us elaborate on this further. In Pilot-PSI the rise time of the magnetic field pulse is $\sim1\,$s,  whereas the first plasma reaches the samples at $\sim200\,$ms.  Subsequently, the plasma density gradually builds up to its maximum value which it attains at the plateau. Our exposures have been complemented by camera observations in several but not all discharges and altogether we have observed $43$ trajectories, that correspond to a small fraction of the detached grains as determined by the post exposure SEM analysis ($\sim5\%$). According to the cameras, all the observed dust remobilization events take place between $0.4$ and $0.7\,$s of plasma exposure. Since the dust surface temperature can only increase in time and the cameras observe the hottest dust\,\cite{hotdust}, we can conclude that after a certain time no dust is released from the samples. This is strongly supported by the results of the re-exposure experiments. Namely, even samples with drastic remobilization activity upon their first exposure, did not exhibit a single incidence of remobilization after re-exposure to an identical discharge. This implies that, as the plasma profiles build up, once the plasma forces become strong enough to satisfy any of the three remobilization conditions for some of the grains, these grains are immediately displaced and further exposure to similar conditions will not lead to any additional activity.

The remobilization experiments were carried out under varying plasma and surface conditions ($n=10^{17}-6\times10^{20}\,$m$^{-3}$, $T_{\mathrm{e}}=0.4-15\,$eV, $T_{\mathrm{i}}=0.4-10\,$eV, $\lambda_{\mathrm{D}}/R_{\mathrm{d}}=10^{-2}-10$, $R_{\mathrm{q}}/\lambda_{\mathrm{D}}=10^{-2}-2$, $\measuredangle\boldsymbol{B}=5^{\circ}-90^{\circ}$). In such a wide range of experimental conditions, we have identified a clear dependence on the dust size. In accordance with the theory, larger grains and agglomerates tend to remobilize much easier than smaller grains $\lesssim10\,\mu$m. Consequently, smaller grains can be expected to have a longer lifetime on PFCs. This is particularly important, in view of recent dust studies in full metal tokamaks which have revealed that W grains with sizes $\lesssim10\,\mu$m are the dominant component of the tungsten dust inventory. For instance, in dust collection activities from the full tungsten ASDEX Upgrade\,\cite{adhesi1,adhesi3} the majority of W dust has sizes less than $10\,\mu$m with a most probable size $\simeq1\,\mu$m, whereas in the first dust studies from JET with the ITER-Like Wall\,\cite{adhesi4} $\sim5\,\mu$m is a typical size of W dust.

We shall now view the theoretical estimates presented in section \ref{theorysection} and the postulations about the three remobilization conditions in light of the experimental results.  Our calculations of the pull-off force strength indicate that it is hard for plasma forces acting along the surface normal to compensate for adhesion and that sliding or rolling can be realized easier than direct lift-up. However, overall, a significant number of grains were detached from the samples during plasma exposure. There are three possible explanations for this discrepancy: \textbf{(i)} Micrometer scale roughness acts as a ramp. While the remobilized dust grains initially slide or roll, they soon encounter an asperity that redistributes the local normal and tangential velocity components causing detachment. \textbf{(ii)} Nanometer scale roughness decreases the ideal strength of the pull-off force by two orders of magnitude. Theoretical models\,\cite{TaborA1,TaborA2} indicate that the influence of nano-roughness on adhesion depends on the value of the dimensionless adhesion parameter $\Theta=(E^*\sigma_{\mathrm{asp}}^{3/2}r_{\mathrm{asp}}^{1/2})/(r_{\mathrm{asp}}\Gamma)$ where $r_{\mathrm{asp}}<a\ll{R}_{\mathrm{d}}$ is the mean asperity radius and $\sigma_{\mathrm{asp}}$ is the standard deviation of asperity heights\,\cite{nanorou}. According to experiments, for large values of the adhesion parameter, $\Theta>10$, the pull-off force can be decreased by at least one order of magnitude\,\cite{TaborA1,TaborA2,TaborA3,TaborA4,TaborA5}. Such high values require that the materials have large Young's moduli $E$. In fact, refractory metals are characterized by large $E$ and especially tungsten has one of the largest values, $E\simeq410\,$GPa. \textbf{(iii)} There are additional plasma induced forces acting on the normal direction apart from the ion drag force and the electrostatic sheath force. We consider this possibility as unlikely to occur and mention it for the sake of completeness.
Some examples are forces due to thermal shocks or $\boldsymbol{J}\times\boldsymbol{B}$ forces. Camera observations in the Pilot-PSI experiments reported here support the first explanation, since they never revealed evidence of a direct lift-up. In the perpendicular configuration all grains moved radially along the sample surface, whereas in the oblique configuration the initial phase of trajectory was either pure tangential or with a dominant tangential component. However, as aforementioned, we did not manage to resolve the starting point of the trajectories. In further support of the first explanation, there were events where the grains were either significantly or slightly displaced from their original position without leaving the sample.

Future dust remobilization experiments will focus on clarifying which of the three remobilization conditions is satisfied first. Recently, we managed to drastically improve the camera resolution in the Pilot-PSI experiments to $9\,\mu$m/pixel\,\cite{Mgrain4}, which can assist in resolving dust trajectories closer to the remobilization instant as well as clarify whether the tangential trajectories observed are along the surface of the sample or along the plasma sheath. Moreover, in experiments with mirror polished samples ($R_{\mathrm{q}}\simeq30\,$nm) micron-roughness can be eliminated. Finally, measurements of the pull-off force either by atomic force microscopy or by electrostatic mobilization can quantify the effect of nano-roughness for W on W adhesion.

A quantitative analytical theory of dust remobilization is currently a formidable task. Self-consistent modelling of remobilization requires a consideration of the influence of dust on the local plasma parameters. Even for a single dust grain residing on the much larger PFC, while the global sheath structure will be unaffected, local effects will still be important owing to the strong dependence of the plasma forces on the micro-flows and on the details of the density profiles (sheath within a sheath case). Theoretical analysis can become even more complex for dust sizes comparable to the Debye length, for multiple grains due to shadowing effects and agglomerates, for roughness values comparable to the Debye length or for castellated PFCs. However, the static nature of the phenomenon implies that particle-in-cell numerical modelling\,\cite{PICmod1,PICmod2} is a viable candidate that can provide quantitative results. Such investigations will be the subject of future work.

In view of such theoretical difficulties, controlled dust remobilization experiments in fusion environments are essential. The experimental technique proposed herein not only realistically mimics naturally occurring sticking impacts, but also allows for an unambiguous quantification of the dust remobilization activity.

\section*{Additional information}

Animations containing alternating SEM images of W samples prior to and post plasma exposure allow for an evaluation of the remobilization activity without the use of overlaid images.
Such animations are provided as supplementary information and available online.

\section*{Acknowledgments}

Controlled pre-adhesion was made possible thanks to the engineering skills of Giambattista Daminelli. The authors would like to thank L. Frassinetti and H. Bergs{\aa}ker for assistance in the experiments carried out on EXTRAP-T2R. P. T. would like to thank H. Bergs{\aa}ker for bringing to his attention the issue of dust accumulation in gaps. This work has been carried out within the framework of the EUROfusion Consortium and has received funding from the European Union’s Horizon 2020 research and innovation programme under grant agreement number 633053 (Enabling Research project CfP-WP14-ER-01/VR-01). The views and opinions expressed herein do not necessarily reflect those of the European Commission.

\end{document}